\documentclass{article}
\usepackage[utf8]{inputenc}
\usepackage[letterpaper]{geometry}
\usepackage[parfill]{parskip}
\usepackage{natbib}
\usepackage{xr-hyper,refcount}
\usepackage{graphicx}
\usepackage[utf8]{inputenc} 
\usepackage[T1]{fontenc}    

\usepackage[hyphens, spaces, obeyspaces]{url}     
\usepackage{hyperref}

\hypersetup{
    colorlinks=true,
    linkcolor=blue,
    filecolor=magenta,      
    urlcolor=blue,
}

\usepackage{booktabs}       
\usepackage{amsfonts}       
\usepackage{nicefrac}       
\usepackage{microtype}      
\usepackage[parfill]{parskip}
\usepackage{algorithm,algorithmic}
\usepackage{amsmath,amsthm,amssymb,bbm}
\usepackage{mathtools}
\usepackage{cases}
\usepackage{comment}
\usepackage{algorithm,algorithmic}
\usepackage{color}
\usepackage{appendix}
\usepackage{xspace}

\usepackage{enumitem}
\newcommand{\hypothesis}[2]{$#1#2$}

\usepackage[english]{babel}
\usepackage{authblk}

\usepackage{libertine}
\usepackage{libertinust1math}
\usepackage[T1]{fontenc}

\usepackage{todonotes}
\usepackage{graphicx}
\usepackage[font=small]{caption}
\usepackage{subcaption}
\usepackage{float}
\usepackage{dirtytalk}
\usepackage{array,multirow,graphicx}

\makeatletter
\newcommand{\printfnsymbol}[1]{\textsuperscript{\@fnsymbol{#1}}}
\makeatother

\DeclareMathOperator*{\HFR}{HFR}

\title{
Unpacking the Drop in COVID-19 Case Fatality Rates:\\
A Study of National \& Florida Line-Level Data
}

\author{Cheng Cheng\thanks{Equal contribution.}, Helen Zhou\printfnsymbol{1}, Jeremy C. Weiss, Zachary C. Lipton\\
Carnegie Mellon University\\
\text{\{\href{mailto:ccheng2@cmu.edu}{ccheng2},
\href{mailto:hlzhou@andrew.cmu.edu}{hlzhou},
\href{mailto:jeremyweiss@cmu.edu}{jeremyweiss},
\href{mailto:zlipton@cmu.edu}{zlipton}\}@cmu.edu}}

\date{August 2020}

\begin{document}
\maketitle

\begin{abstract}
    Since the COVID-19 pandemic first reached the United States, 
the case fatality rate has fallen precipitously.
In mid-April, when cases first peaked, 
the case fatality rate was 
$7.9\%$. 
By the second peak in mid-July, 
the rate dropped to
the $1\%$--$2.3\%$ range, 
where it has remained ever since. 
In both academic articles 
and the broader public discourse,
several possible explanations 
have been floated,
including greater detection of mild cases
due to expanded testing,
shifts in age distribution 
among the infected, 
lags between confirmed cases and reported deaths, 
improvements in treatment,
mutations in the virus, 
and decreased viral load as a result of mask-wearing. 
Using both Florida line-level data 
and recently released (but incomplete)
national line level data 
from April 1, 2020 to November 1, 2020
on cases, hospitalizations, 
and deaths---each stratified by age---we 
unpack the drop in case fatality rate (CFR). 
Under the hypothesis that improvements in treatment efficacy 
should correspond to decreases in hospitalization fatality rate (HFR),
we find that improvements in the national data 
do not always match the story told by Florida data.
In the national data, treatment improvements
between the first wave and the second wave 
appear substantial 
(with the relative decrease in HFR as little as $24\%$ in the 80+ age group 
and as much as $50\%$ in the 30-39 age group), 
but modest when compared to the drop in aggregate CFR 
(in the range of $70.9\%$--$87.3\%$). 
By contrast, possibly due to constrained resources 
in a much larger second peak, 
Florida data suggests comparatively little difference 
between the first and second wave, 
with HFR slightly \emph{increasing} in every age group 
(by as little as $2\%$ in the $80+$ age group 
and as much as $12\%$ in the 60-69 age group). 
However, by November 1st, both Florida and national data 
suggest significant decreases in HFR since April 1st---at 
least $42\%$ in Florida and at least $45\%$ nationally in every age group.
By accounting for several confounding factors, 
our analysis shows how age-stratified HFR
can provide a more realistic picture 
of treatment improvements than CFR. 
One key limitation of our analysis is that 
the national line-level data remains incomplete
(covering roughly one third of national cases)
and plagued by artifacts
(some states' data have all cases filed onto one of a few dates).
Our analysis highlights the crucial role that this data can play
but also the pressing need 
for public, complete, and high-quality
age-stratified line-level data 
for both cases, hospitalizations, and deaths for all states. 
 
\end{abstract}

\section{Introduction}
\label{sec:intro}
Over the past year, 
the coronavirus disease (COVID-19) pandemic has continually evolved, 
with disease outbreaks expanding and contracting
(and expanding yet again); 
lockdown measures tightening and loosening;
testing capacity increasing
(and occasionally decreasing,
due to production shortages~\citep{wu2020groundhog});
and treatments protocols evolving.
With lives and livelihoods in the balance,
public officials, clinicians, and business leaders
have tried to maintain a grasp of the
the rapidly unfolding situation,
looking to the publicly reported aggregate data
to ask the key questions that could guide policy decisions:

Have new treatment protocols improved outcomes over time?
Is COVID-19 fatality decreasing overall?
How does the infection rate today compare to at previous dates
when alternative lock-down protocols were in place?
To what extent are the rising confirmed case numbers 
seen during the second peak
attributable to increased infections 
versus to higher detection rates 
due to increased testing capacity?

Depending on the answers to these questions, 
local officials might consider changing lockdown measures, 
hospitals may need to allocate more resources, 
and business leaders might decide 
to adjust corporate policies and services.

\begin{figure}
\includegraphics[width=\textwidth]{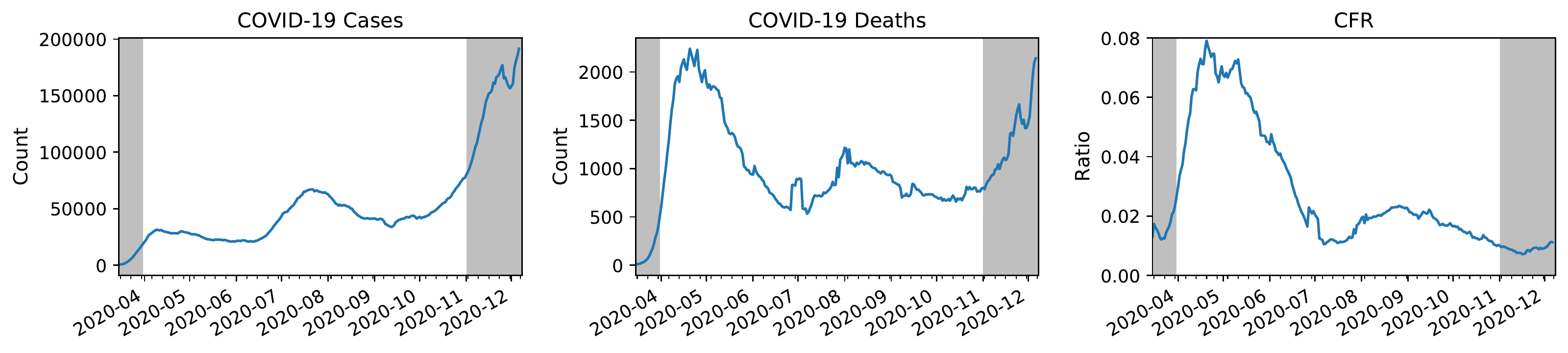}
\caption[cap]{From left to right: confirmed cases, 
deaths, and case fatality rate, 
calculated using 7-day trailing averages \hspace{0.1em} 
based on national reporting data available 
via USAFacts \citep{usafacts} and pulled from 
the Carnegie Mellon Delphi project's COVIDcast API \citep{covidcast2020api}. Data outside the April 1st to November 1st time range considered in this study is grayed out.}
\label{fig:national-cfr}
\end{figure}

Consider the two most widely reported statistics,
confirmed cases and confirmed deaths.  
At first glance, the two appear 
to tell radically different stories
about the trajectory of the pandemic over time.
Confirmed cases appear 
to have first peaked in early April
with a 7-day average of nearly $32,000$ daily cases,
followed by a much larger second wave
with 7-day averages peaking at nearly $67,000$ daily cases 
(Figure \ref{fig:national-cfr}, left panel).
However, reported deaths appear to tell 
a contradictory story concerning 
the relative severity of the two waves 
(Figure \ref{fig:national-cfr}, middle panel).
Overall the case fatality rate, 
calculated by dividing 7-day averaged COVID-19 confirmed deaths
by confirmed cases on each day,
appears to have fallen dramatically 
between the first wave and second wave, 
from nearly 7.9\% at the height
of the first wave in mid-April 
to the 1\%--2.5\% range in mid-July, 
where it has been comparatively stable 
for the last two months 
(Figure \ref{fig:national-cfr}, right panel)\footnote{These numbers come from USAFacts \citep{usafacts} through the COVIDcast API \citep{covidcast2020api}, however we arrive at similar numbers when using the national CDC data \citep{cdc_data}}.

In this paper, we set out to answer the question that immediately follows:
\emph{what explains the movement (and apparent overall decline)
in case fatality rate over the course of the COVID-19 pandemic?}
There are several plausible explanations,
each with significant policy implications for stakeholders.
So far, the public discourse appears 
to center around following hypotheses:

\begin{enumerate}[label=(\hypothesis{H}{{\arabic*}})]
    \item The age distribution among infected patients has shifted,
            altering the fatality rate significantly due to the
            comparative higher risk among the geriatric populations \citep{thomp2020,moser2020changing, horwitz2020trends}.
    \item Testing capacity has gone up significantly,
            with case fatality rate driven down primarily
            due to the rising number of tests performed 
            \citep{fan2020decreased, mad2020}.
    \item Apparent shifts in case fatality rate,
            are artifacts due to the delay between
            detection and fatality \citep{thomp2020, madrigal2020second}. 
    \item Treatment has improved as doctors grow more experienced
            and new therapeutics become available, 
            lowering the fatality rate over time 
            \citep{levy2020gates, horwitz2020trends, beigel2020remdesivir, recovery2020dexamethasone,self2020effect}. 
    \item The disease itself is mutating,
            leading to changes in the actual infection fatality rate 
            over time \citep{pachetti2020impact,fan2020decreased}. 
    \item Social distancing precautions have reduced the viral load 
    that individuals are exposed to,            
    resulting in less severe infections 
    \citep{eldeclining, pachetti2020impact, piubelli2020overall}.  
\end{enumerate}

Note that when the age distribution shifts 
to a younger demographic (H1), 
the dropping aggregate case fatality rate can be misleading 
due to differences in fatality between different age groups. 
Additionally, the next two phenomena---testing 
ramp-up and delays between detection (H2) and fatality (H3)---can 
cause the behavior of case fatality rate to diverge substantially 
from the behavior of the true infection fatality rate. 
Thus, if policy-makers use the aggregate case fatality rate 
to represent the severity of COVID-19, 
this could result in misinformed policy decisions.

On the other hand, the last three phenomena---improved 
treatments, disease mutation, and changing viral load---impact 
both the case fatality rate and the actual infection fatality rate, 
and could be reasonable grounds for policy considerations such as re-opening.
In this paper, \textbf{we demonstrate 
how given accurate, sufficiently granular data,
the first three phenomena (``artifacts") 
can be accounted for when attempting 
to quantify improvements in treatment} (H4).
We note, however, that without additional data,
H5 and H6 cannot be decisively separated out from H4.

In particular, 
we argue that complete and accurate 
\emph{age-stratified hospitalization data}
are pivotal for distinguishing true improvements from artifacts.
Hospitalizations should be less influenced 
by testing capacity than confirmed cases, 
and less influenced by treatment efficacy than deaths. 
Additionally, compared to the general population, 
testing among the inpatient population has been relatively thorough
(compared to the general population)
throughout the course of the pandemic.
While there may have been changes 
in admitting criteria at the very worst moments
\citep{phua_icu_manage}\citep{uptodate_eval},
for example, when New York hospital demand 
exceeded capacity in late March, for the most part, 
criteria for inpatient hospitalization 
is relatively consistent across time periods. 
Regrettably, however, reliable line-level 
age-stratified hospitalization data 
is not yet publicly available 
for most states \citep{propublica_hosp, data_problems_nyt}, 
leaving central questions unresolved. 

We center our analysis on (1) state-level data made available 
by the Florida Department of Health (FDOH) \citep{florida_line_data} 
and (2) incomplete national-level data made available 
by the Centers for Disease Control and Prevention (CDC) \citep{cdc_data}.
Both the FDOH and CDC have demonstrated an uncommon openness 
by sharing line-level data, including date of detection,
demographics including age and gender, 
and indicators of eventual hospitalization and death.
The line-level nature of the data allows us 
to perform a cohort-based analysis,
generating descriptive statistics 
comparing case confirmations, hospitalizations, and deaths,
among cohorts of patients defined 
by the date on which their infection was detected.
By contrast, reported case fatality rates are typically not cohort-based---the
patients whose deaths are reported in the numerator 
are not in general the same patients 
whose confirmed infections show up in the denominator.
Because case confirmation tends to precede reported deaths,
these signals tend to be misaligned
and are subject to fluctuation,
even if the actual case fatality rate were fixed
(so long as incidence does change). 
Line-level data enable us to circumvent this problem. 
Moreover, the availability of age and gender data
allows us both to track demographic shifts over time,
and to perform age-stratified analyses of fatality rates.

Importantly, this analysis yields several important observations:
(i) testing increased between the first and second waves, 
but does not explain away these waves; 
(ii) since age distributions shifted substantially 
between the first and second waves, 
age must be accounted for in order to separate out 
the effects of treatment from age shift;
(iii) age-stratified hospitalization fatality rates improved substantially 
between the first and second wave in the national data (with HFR decreasing by as little as $24\%$ in the 80+ age group 
and as much as $50\%$ in the 30-39 age group), 
but by contrast were relatively unchanged between the first and second wave in Florida (with a slight \emph{increase} in HFR by as little as $2\%$ in the $80+$ age group 
and as much as $12\%$ in the 60-69 age group);
(iv) nationally, the relative drop in HFR appears to be negatively associated with age (improvements in fatality rate have disproportionately benefited the young);
(v) by November 1st, both Florida and national data 
suggest significant decreases in HFR since April 1st---at 
least $42\%$ in Florida and at least $45\%$ nationally in every age group;
and (vi) comprehensive age-stratified hospitalization data 
is of central importance 
to providing situational awareness 
during the COVID-19 pandemic, 
and its lack of availability 
among public sources for most states 
(and the extreme incompleteness of national data)
constitutes a major obstacle to tracking and planning efforts.

\subsection{Related Work}
\label{sec:related_work}
Throughout the course of the pandemic, 
several treatments have been proposed,
with randomized controlled trials 
designed to test for their effectivenesss. 
Dexamethosone, a corticosteriod 
commonly prescribed for other indications,
resulted in a lower 28-day mortality 
among patients hospitalized with COVID-19 
and receiving respiratory support 
\citep{recovery2020dexamethasone}. 
Among adults hospitalized with COVID-19 
who had evidence of lower respiratory tract infection, 
broad-spectrum antiviral remedisivir 
was associated with shortened time to recovery
\citep{beigel2020remdesivir, madsen2020remdesivir}.
Clinical trials for hydroxychloroquine 
\citep{self2020effect, horby2020effect}
and convalescent plasma \citep{agarwal2020convalescent}
found no positive results in prevention 
of further disease progression or mortality. 
Recently in November,
(outside our study time period), 
monoclonal antibody treatments bamlanivimab and the combination therapy casirivimab and imdevimab
were approved for emergency use authorization \citep{monoclonal_eua, regeneron_monoclonal_eua}. 
Unlike dexamethosone and remdesivir, 
these therapies are not recommended for hospitalized patients \citep{dyer2020covid},
but instead have been shown to have greatest benefit in unadmitted COVID-19 patients 
likely to progress to severe COVID-19 (for bamlanivimab) \citep{chen2020sars}, 
and in patients who have not yet mounted their own immune response or who have high viral load (for casirivimab and imdevimab) \citep{regeneron_press}.
While these clinical trials have evaluated 
the effects of specific treatments 
in their identified target populations, 
we are interested in the broader impacts 
of treatment improvements over time 
as they have been used in practice at a larger scale.

One way to get a holistic sense 
of improvements over time 
is by examining fatality rates.
In a study of 53 countries, 
all but ten were found to have 
lower case fatality rates in the second wave 
compared to the first \citep{fan2020decreased}.
In a study conducted among patients in England 
admitted to critical care between March 1st until May 30th, 
it was found that after adjusting for age, sex, ethnicity, 
comorbidities, and geographic region, 
mortality risk in mid-April and May 
was markedly lower compared to earlier in the pandemic. 
Among hospitalized COVID-19 patients 
in a single health system in New York City, 
\citet{horwitz2020trends} demonstrate 
that after adjusting for age, sex, ethnicity, 
and several clinical factors, 
mortality between March 1st and June 20th 
decreased but not as much as observed 
before adjusting for these factors.
We note that while these studies 
provide thorough estimates of mortality 
for their respective regions 
and for their specific time periods, 
we analyze data that purportedly captures
all of Florida and most of the United States, 
and employ a method which allows us to estimate the trend 
between any pair of dates without re-fitting. 

While our data does not contain sufficient information 
about viral mutation patterns and social distancing 
resulting in reduced viral loads (H5 and H6), 
there has been limited work investigating their impact. 
Among seven different countries,
\citet{pachetti2020impact} quantify the drop in mortality, 
and find a correlation between declining CFR 
and employing strict lockdown policies 
as well as widespread PCR testing. 
At a hospital system in northern Italy, 
\citet{piubelli2020overall} found 
that among patients diagnosed with COVID-19 in their emergency room, 
the proportion of patients requiring intensive care decreased over time, 
along decreasing median values of viral load. 
Our analysis does not attempt to separate out 
the effects of viral mutations (H5) 
and changing viral loads (H6),
but we note that these are all factors 
that can affect change in the true infection fatality rate,
and therefore can be reflected in our estimates as well.

\section{Methodology}
\label{sec:methods}
\subsection{Data}
The two main data sources that we use 
are the Florida COVID-19 Case Line data 
\citep{florida_line_data} released by (FDOH), 
and the national COVID-19 Case Surveillance 
Data \citep{cdc_data} released by CDC. 

\vspace{0.5em}\noindent\textbf{Cohort Selection}~
All cases from Florida are COVID-19 cases 
confirmed with a PCR-positive lab result.
In order to conduct comparable analyses, 
we also filter the national data to cases 
that are confirmed with a positive PCR result. 
We filter for only the cases identified 
between March 26th, 2020 and Novermber 1st, 2020 
to ensure that at the time of our analysis, 
each patient has had at least 30 days 
to have their hospitalization or death recorded
from their initial case confirmation date (Florida) 
or CDC report date (national). 
Since the CDC data is released on a monthly basis,
this is the widest time range of data available 
at the time of our analysis.
For the national CDC data,
three states (NJ, IL, and CT) 
appear to have all of their cases 
reported on one or two dates 
(Appendix Figure \ref{fig:exclude_states}). 
Therefore, we removed them from our analysis.
     
\vspace{0.5em}\noindent\textbf{Pre-processing}~
Each case is labeled with whether the patient 
was eventually hospitalized or deceased. 
These labels take on four categories: 
\say{yes}, \say{no}, \say{unknown} and \say{missing} (Table \ref{tab:hosp_death}).
In both the Florida and national data, 
unknown corresponds to checking an ``unknown" box,
whereas missing corresponds to leaving the question empty. 
Note that $42.5\%$ of the Florida cases 
have unknown hospitalization data,
and $46.3\%$ of the national cases 
have unknown or missing hospitalization data. 
For our analyses, we coded the \say{unknown} 
and \say{missing} categories into \say{no}.  

\begin{table*}[thb] 
\centering
\caption[cap]{Original categories for hospitalization 
and death in Florida and national data}.
  \setlength{\tabcolsep}{1em}
  \resizebox*{0.9\textwidth}{!}{

\begin{tabular}{lll|ll}
\toprule
{} & \multicolumn{2}{c}{Florida} & \multicolumn{2}{c}{Country} \\
\midrule
{} & Hospitalization &           Death &  Hospitalization &            Death \\
Yes     &    50414 (6.2\%) &    18333 (2.3\%) &    369377 (6.6\%) &    126994 (2.3\%) \\
No      &  425985 (52.8\%) &        0 (0.0\%) &  2551008 (45.8\%) &  2631070 (47.3\%) \\
Unknown &  327622 (40.6\%) &        0 (0.0\%) &   691973 (12.4\%) &   588668 (10.6\%) \\
Missing &     2688 (0.3\%) &  788376 (97.7\%) &  1953812 (35.1\%) &  2219438 (39.9\%) \\
\bottomrule
\end{tabular}}
  \label{tab:hosp_death}
\end{table*}

\vspace{0.5em}\noindent\textbf{Demographics}~
The demographics of the Florida 
and national cohorts can be found in Table \ref{tab:demographics}.
\begin{table*}[thb] 
\centering
\caption[cap]{Demographics of the Florida and national cohorts}.
  \setlength{\tabcolsep}{1em}
  \resizebox*{0.7\textwidth}{!}{
    \begin{tabular}{l|llll}
 \toprule
 & \hspace{5mm}{} &         Florida &          Country \\
 \midrule
 \parbox[t]{2mm}{\multirow{17}{*}{\rotatebox[origin=c]{90}{Demographics}}}  
& Lab Confirmed COVID-19 Cases &          806709 &          5566170 \\
  & Age            &                 &                  \\
  & \hspace{5mm}0-9            &    28763 (3.6\%) &    188885 (3.4\%) \\
  & \hspace{5mm}10-19          &    72176 (8.9\%) &    540922 (9.7\%) \\
  & \hspace{5mm}20-29          &  155424 (19.3\%) &  1106791 (19.9\%) \\
  & \hspace{5mm}30-39          &  138599 (17.2\%) &   918351 (16.5\%) \\
  & \hspace{5mm}40-49          &  124778 (15.5\%) &   837394 (15.0\%) \\
  & \hspace{5mm}50-59          &  120581 (14.9\%) &   800695 (14.4\%) \\
  & \hspace{5mm}60-69          &   81593 (10.1\%) &   558180 (10.0\%) \\
  & \hspace{5mm}70-79          &    48223 (6.0\%) &    318363 (5.7\%) \\
  & \hspace{5mm}80+            &    35531 (4.4\%) &    257045 (4.6\%) \\
  & \hspace{5mm}Unknown       &     1041 (0.1\%) &     39544 (0.7\%) \\
  & Gender         &                 &                  \\
  & \hspace{5mm}Female         &  415089 (51.5\%) &  2842355 (51.1\%) \\
  & \hspace{5mm}Male           &  388103 (48.1\%) &  2611573 (46.9\%) \\
  & \hspace{5mm}Missing        &        0 (0.0\%) &     68304 (1.2\%) \\
  & \hspace{5mm}Unknown        &     3517 (0.4\%) &     43938 (0.8\%) \\
  \midrule
  \parbox[t]{2mm}{\multirow{6}{*}{\rotatebox[origin=c]{90}{Outcome}}} 
  & Hospitalized   &                 &                  \\
  & \hspace{5mm}Yes            &    50414 (6.2\%) &    369377 (6.6\%) \\
  & \hspace{5mm}No             &  756295 (93.8\%) &  5196793 (93.4\%) \\
  & Died           &                 &                  \\
  & \hspace{5mm}Yes            &    18333 (2.3\%) &    126994 (2.3\%) \\
  & \hspace{5mm}No             &  788376 (97.7\%) &  5439176 (97.7\%) \\
 \bottomrule
\end{tabular}
 }
  \label{tab:demographics}
\end{table*}

\subsubsection{Secondary Data}
To supplement our analysis, we use two secondary data sources:
(1) COVID-19 testing data from COVID Tracking Project \citep{covid_tracking_project}; 
and (2) COVID-19 confirmed cases and deaths from USAFacts \citep{usafacts} 
and pulled from the Carnegie Mellon Delphi project's 
COVIDcast API \citep{covidcast2020api}. 
In contrast to our line-level data, 
these two data sources provide incidence data 
(e.g. new deaths that day). 
We visualize this data across a larger time period 
between March 9th and December 1st, 
but gray out the period outside 
of our study's time range 
(between April 1st and November 1st). 

\subsection{Signal Smoothing}
To view the progression of the pandemic over time with reduced noise, 
for each date, we compute the 7-day lagged average 
for COVID-19 cases, hospitalizations, deaths,
and tests for Florida and the nation. 
From this point in the paper on,
whenever we directly use COVID-19 cases, 
hospitalizations, deaths and tests or calculate CFR, HFR, 
and positive test rates based on them, 
unless otherwise stated, 
we are referring to the smoothed signal.
For the Florida FDOH and national CDC data,
we collect data extending back to March 25th 
in order to conduct our analyses 
on the April 1st to December 1st time range.
For the two secondary data sources,
we collect data extending back to March 9th 
in order to visualize the data starting from March 15th.

\subsection{Separating Artifacts from True Improvements}
We argue that three main phenomena 
fuel a dramatic ``artificial" decrease in CFR: 
increased testing capacity (H1), 
shifting age distributions (H2), 
and delays between detection and fatality (H3). 

To establish the phenomenon of increased testing capacity, 
we visualize 7-day lagged averages of testing capacity 
and the proportion of tests returning positive 
in Florida and the United States 
using data from The COVID Tracking Project \citep{covid_tracking_project}. 
To avoid artifacts from increased testing (H1),
when considering treatment improvements, 
we examine changes in \emph{hospitalization} fatality rates 
rather than case fatality rates. 

To establish and account for shifting age distributions (H2),
we examine cases, hospitalizations, and deaths 
stratified by age groups: 
0-9, 10-19, 20-29, 30-39, 40-49, 
50-59, 60-69, 70-79, and 80+. 
Naturally, age stratification reduces the amount of data 
that goes into each estimate, 
so when computing estimates of HFR 
we omit results that are based on fewer than two deaths.

Finally, to account for delays 
between detection and fatality (H3),
we take advantage of line-level data for each individual. 
For each date, we extract the cohort of individuals 
confirmed positive on that date, 
as well as whether those individuals 
\emph{eventually} died or were hospitalized. 
To provide enough time for patients' 
eventual hospitalizations or deaths to be recorded, 
we filter out rows with positive specimen dates 
within 30 days of the last time the data was updated.
Since CDC data was last updated on December 4th, 
we only use data from November 1st or earlier. 
Florida data is updated daily, 
but we use the same time range 
as the CDC data in order to make the plots comparable.

Taking the above three adjustments into account, 
our primary quantity of interest for treatment improvements 
is the age-stratified HFR. 
For the rest of the paper, 
we define CFR and HFR at day $t$ as follows:
\begin{align*}
    \text{CFR}_t &= \frac{\text{cases confirmed (or reported) at day $t$ that eventually die}}{\text{cases confirmed (or reported) at day $t$}}\\
    \text{HFR}_t &= \frac{\text{cases confirmed (or reported) at day $t$ that eventually get hospitalized and die}}{\text{cases confirmed (or reported) at day $t$ that eventually get hospitalized}}
\end{align*}
Here, ``eventually" means that the hospitalization or death 
was recorded by the date of data collection (i.e. December 4th), 
which gives each patient at least 30 days 
after their case confirmation/report date 
to have the events recorded.

\subsection{Quantifying True Improvements}
Thus far, news and academic sources have highlighted 
three main ``true improvements": 
improvements in treatment (H4), 
disease mutations (H5), 
and reduced viral loads due to social distancing (H6). 
We seek to quantify treatment improvements (H4) 
by computing the decrease in hospitalization fatality rate.
Although practice is constantly evolving,
major improvements in treatment in our study time range 
such as dexamethasone \citep{recovery2020dexamethasone} 
and remdesivir \citep{beigel2020remdesivir} 
have primarily targeted hospitalized patients, 
and we expect that improvements 
due to those treatments should be reflected in the HFR. 

In order to quantify the change in HFR with uncertainty,
we use a block-bootstrapping technique with post-blackening \citep{bootstrap_davison_hinkley_1997}.
This involves fitting cubic smoothing splines 
to each age group's 7-day averaged HFRs, 
computing the residuals, block-bootstrap
resampling $1000$ replicates of the residuals, 
and adding the residuals back onto the fitted splines 
in order to create $1000$ replicates 
sampled from the original data distribution. 
By re-fitting cubic smoothing splines to each of these datasets, 
we can estimate each day's HFR with $95\%$ confidence intervals. 
See Appendix \ref{app:block_bootstrap} 
for details of the procedure applied to our data.

\section{Results}
\subsection{Cases, Hospitalizations, and Deaths}
\label{sec:counts}
\begin{figure}[ht!]
\centering
\begin{subfigure}[t]{0.95\textwidth}
\includegraphics[width=\textwidth]{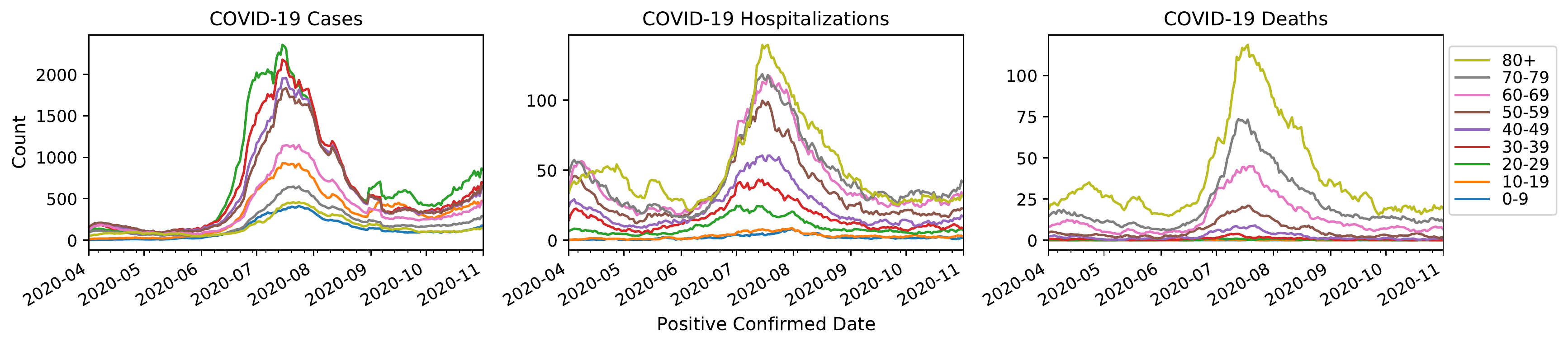}
\caption{Florida FDOH Data}
\label{fig:covid_trend_age}
\end{subfigure}
\begin{subfigure}[t]{0.95\textwidth}
\vspace{0.8em}
\includegraphics[width=\textwidth]{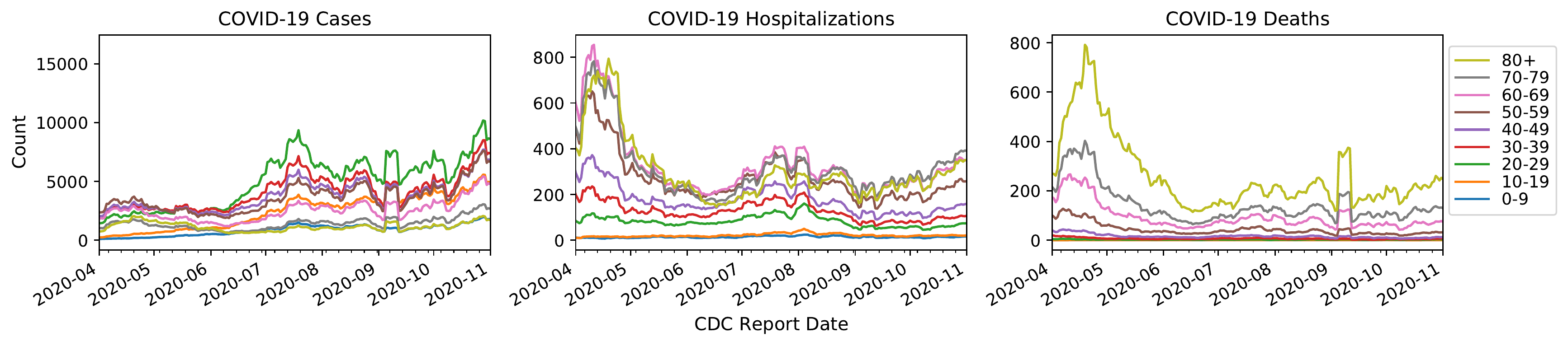}
\caption{United States CDC Data}
\label{fig:country_covid_trend_age}
\end{subfigure}
\caption{Age-stratified cases, (eventual) deaths, 
and (eventual) hospitalizations in Florida and in the United States,
by the date of first positive test result (Florida) 
and date of report to the CDC (U.S.), respectively.
Note that the $x$ axis is 
\emph{not}
the date of death or date of hospitalization.
}
\label{fig:covid_trend}
\end{figure}

In both the FDOH and the CDC datasets,
one can discern two waves of COVID-19 cases, 
the first occurring mid-April and 
the second occurring mid-July (Figure \ref{fig:covid_trend}).
Cases appear to be on the rise leading up to November,
however, they have not yet reached a peak 
within the study time range. 
Overall, we find strong evidence 
for peaks in COVID-19 infections in mid-April and mid-July, 
with Florida undergoing a more severe second peak
than its first peak 
(Appendix Figure \ref{fig:covid_trend_fl}), 
whereas nationally in aggregate the second peak 
appears less severe than the first peak 
(Appendix Figure \ref{fig:country_covid_trend_agg}).

\begin{figure}[ht!]
\centering
\includegraphics[width=\textwidth]{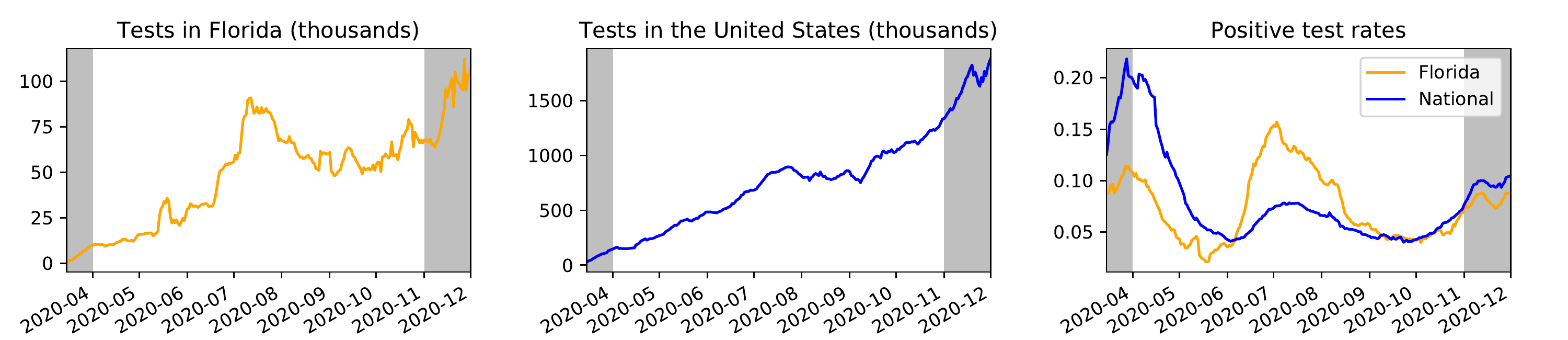}
\caption{COVID-19 positive test rates (right) 
and amount of testing (left and middle) 
for Florida and the United States as a whole, 
calculated using 7-day trailing averages 
and pulled from the COVID Tracking Project 
\citep{covid_tracking_project}. 
Positive test rate is calculated 
by dividing new positives by total new tests on each day. 
Data outside the April 1st to November 1st time range 
considered in this study is grayed out.}
\label{fig:pos_test_rate}
\end{figure}

\paragraph{Increased Testing} 
While testing increased by $696\%$ in Florida 
and by $435\%$ in the country 
between the first and second waves 
(Figure \ref{fig:pos_test_rate}, left and middle panel), 
our data shows that the increase in testing cannot 
fully account for the second peak. 
Despite increased testing inflating the raw number of cases, 
we still observe two peaks in positive test rates 
in April and July (Figure \ref{fig:pos_test_rate}, right panel). 
In Florida, the second peak is larger than the first, 
whereas nationally the second peak 
is smaller than the first.
Leading up to November, 
positive test rates have been rising 
both in Florida and nationally.

\paragraph{Cases} Across all age strata,
Florida's second peak is much more severe 
than the first peak (Figure \ref{fig:covid_trend_age}, left panel), 
in contrast to the differences 
between the two peaks of the national data 
(Figure \ref{fig:country_covid_trend_age}, left panel). 
In aggregate, Florida cases have risen by $943\%$ 
(Appendix Figure \ref{fig:covid_trend_fl}, left panel),
whereas national cases have increased by $97\%$
(Appendix Figure \ref{fig:country_covid_trend_age}, left panel). 
Note that the national aggregate includes data 
from populous states hard-hit 
in the first wave (e.g. New York),
so it is no surprise that the national picture 
is markedly different from that of Florida. 
Towards November, there is a rise in cases 
both in Florida and nationally.

\paragraph{Hospitalizations and Deaths} 
Overall, hospitalizations and deaths 
corroborate the story told by positive test rate.
In Florida, hospitalizations and deaths again 
indicate a more severe second peak than first peak 
(Figure \ref{fig:covid_trend_age}, center and right panels), 
though the contrast in peak size 
is not as dramatic as in the plot of cases. 
In the national data, the second peak 
is actually \emph{smaller} than the first peak 
(Figure \ref{fig:country_covid_trend_age}, center and right panels). 
Both of these discrepancies of trends 
seen in cases versus in hospitalizations and deaths 
are likely attributable to increases in testing 
(Figure \ref{fig:pos_test_rate}).

\begin{figure}[ht!]
\centering
\begin{subfigure}[t]{0.95\textwidth}
\includegraphics[width=\textwidth]{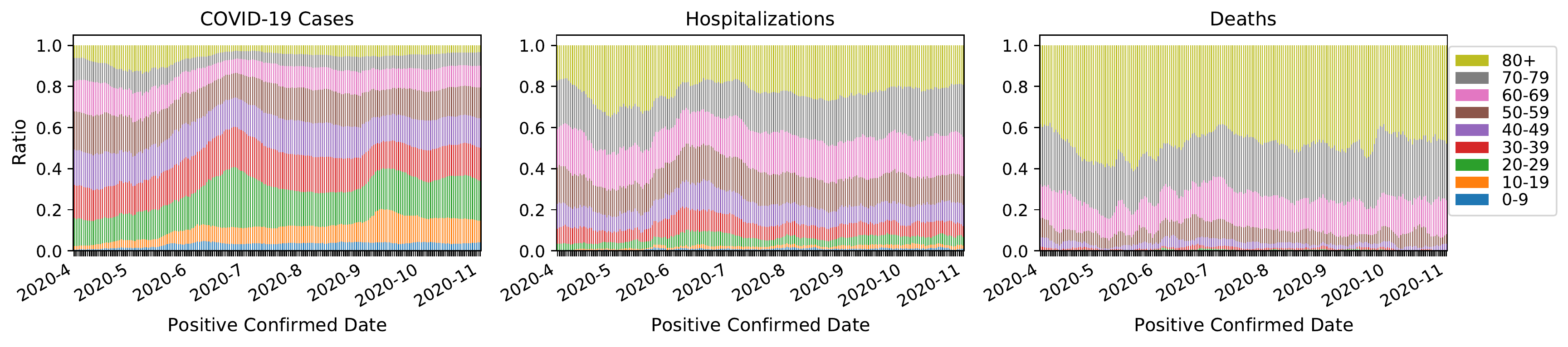}
\caption{Florida FDOH Data}
\label{fig:florida_age_ratio}
\end{subfigure}
\begin{subfigure}[t]{0.95\textwidth}
\vspace{0.8em}
\includegraphics[width=\textwidth]{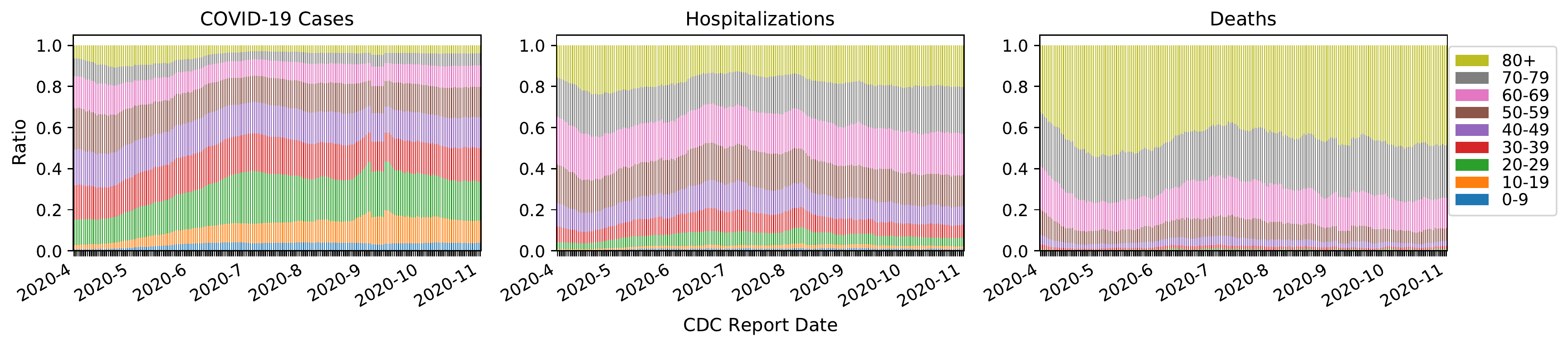}
\caption{United States CDC Data}
\label{fig:country_age_ratio}
\end{subfigure}
\caption{Age distributions among Florida and national cases, 
(eventual) hospitalizations, and (eventual) deaths, 
by the date of first positive test result (Florida) 
and date of report to the CDC (U.S.), respectively.}
\label{fig:covid_ages}
\end{figure}

\subsection{Confounding Due to Demographic Shift}
\label{sec:demographics}
\paragraph{Age} Between the two peaks, 
the age distribution of cases shifted substantially, 
with the median age in Florida changing from 52 to 40,
and the median age group in national data falling from 50-59 to 30-39. 
Since the second peak, the age distributions of cases, 
hospitalizations, and deaths have continued 
to fluctuate (Figure \ref{fig:covid_ages}),
with older individuals comprising a disproportionate share 
of the hospitalization and death counts (Figure \ref{fig:covid_ages}, 
middle and right panel). 

\paragraph{Gender} 
While age appears to vary substantially over time, 
the gender ratios in each age group's cases, 
hospitalizations, and deaths appear relatively flat over time 
(Appendix Figure \ref{fig:covid_gender}). 
Thus, in our plots we choose not to stratify 
by gender 
due to the reasonably small shift 
in the gender distribution over time, 
and practically to have enough support in each group.

\subsection{Age-stratified HFR}
\label{sec:fatality}
Consistent with findings that increased age is associated with higher mortality rates \citep{Mahasem1327_agedeathrate}, we observe that as the age of the group increases, the corresponding HFR increases (Tables \ref{tab:hfr_drop_peak} and \ref{tab:hfr_drop_overall}). Measuring treatment improvements by HFR drop (computed as $\frac{\HFR_{new} - \HFR_{old}}{\HFR_{old}}$), we also observe that in the national data, treatment improvements between the peaks become smaller with increased age. 

Between the two peaks (April 15th to July 15th, Table \ref{tab:hfr_drop_peak}), 
the national age-stratified HFR estimates 
from block bootstrapping decreased by as little as $24\%$ in the 80+ age group,
and as much as $50\%$ in the 30-39 age group.
On the other hand, in Florida the age-stratified HFR 
actually \emph{increased} in each age group 
by as little as $2\%$ in the 80+ age group, 
and as much as $12\%$ in the 60-69 age group.
Note that the HFR changes between peak dates in Florida
are an example of Simpson's paradox, 
where in each age group the HFR increased, 
but the aggregate HFR actually decreased by $2.6\%$.

Compared to peak-to-peak changes, 
changes across the entire time range 
(April 1st to November 1st, Table \ref{tab:hfr_drop_overall}) 
show a more dramatic decrease. 
In Florida, the HFR drops by as little as $42\%$ in the 80+ age range, 
and as much as $53\%$ in the 70-79 age range. 
Nationally, the HFR drops by as little 
as $45\%$ in the 80+ age groups,
and as much as $73\%$ in the 40-49 age group.
As the age group gets older, 
we again see an increase in age-stratified HFR 
and smaller treatment improvements as measured by HFR drop.

While we focus on the two peaks and 
the endpoints of the study time range, 
we also include plots of HFR estimates 
with uncertainty for all dates between 
April 1st and November 1st in Appendix \ref{app:block_bootstrap}.
Consistent with our point estimates, 
the overall HFR in Florida appears relatively flat 
until August, in which the HFR decreases 
greatly across all age groups.
In the national data, there appears to be 
an almost monotonic decline in HFR 
across all age groups for the entire time range,
with the decrease tapering out in August.

When stratifying by gender in addition to age, 
the conclusions surrounding drops in HFR 
are similar to those when just stratifying by age 
(see Appendix \ref{app:gender_hfr_drop}). 

\begin{table}[H]
      \setlength{\tabcolsep}{0.5em}
  \resizebox{0.8\textwidth}{!}{\begin{minipage}{\textwidth}
    \centering

\begin{tabular}{l|ccc|ccc}
\toprule
& \multicolumn{3}{c|}{\textbf{Florida}} & \multicolumn{3}{c}{\textbf{National}} \\
\hline
\textbf{Age group}& 2020-04-15 &            2020-07-15 &        04-15 to 07-15 &            2020-04-15 &               2020-07-15 &        04-15 to 07-15 \\
\midrule
aggregate &         0.24 (0.21, 0.26) &     0.23 (0.22, 0.24) &   -0.026 (-0.14, 0.1) &      0.3 (0.29, 0.31) &        0.18 (0.17, 0.19) &  -0.39 (-0.43, -0.35) \\
30-39     &      - &  - &    - &   0.055 (0.051, 0.06) &      0.027 (0.025, 0.03) &   -0.5 (-0.57, -0.44) \\
40-49     &       - &  - &    - &    0.099 (0.095, 0.1) &     0.056 (0.054, 0.059) &  -0.43 (-0.47, -0.39) \\
50-59     &       0.092 (0.078, 0.11) &     0.1 (0.093, 0.11) &    0.11 (-0.11, 0.37) &     0.17 (0.16, 0.17) &        0.1 (0.099, 0.11) &  -0.37 (-0.41, -0.34) \\
60-69     &         0.19 (0.16, 0.21) &     0.21 (0.19, 0.22) &   0.12 (-0.042, 0.37) &     0.27 (0.26, 0.28) &         0.19 (0.18, 0.2) &  -0.29 (-0.33, -0.26) \\
70-79     &         0.32 (0.29, 0.34) &     0.33 (0.31, 0.34) &  0.035 (-0.067, 0.17) &      0.4 (0.39, 0.42) &         0.29 (0.28, 0.3) &  -0.28 (-0.31, -0.25) \\
80+       &          0.47 (0.44, 0.5) &      0.48 (0.46, 0.5) &    0.02 (-0.053, 0.1) &     0.57 (0.55, 0.58) &        0.43 (0.43, 0.44) &  -0.24 (-0.26, -0.21) \\
\bottomrule
\end{tabular}
  \end{minipage}}
\caption{Estimates of HFR and drop in HFR on peak dates.
Median and $95\%$ confidence intervals are computed using block bootstrapping. Results with inadequate support are omitted.}
    \label{tab:hfr_drop_peak}
\end{table}

\begin{table}[H]
      \setlength{\tabcolsep}{0.5em}
  \resizebox{0.8\textwidth}{!}{\begin{minipage}{\textwidth}
    \centering
\begin{tabular}{l|ccc|ccc}
\toprule
& \multicolumn{3}{c|}{\textbf{Florida}} & \multicolumn{3}{c}{\textbf{National}} \\
\hline
\textbf{Age group}&                2020-04-01 &               2020-11-01 &        04-01 to 11-01 &            2020-04-01 &                2020-11-01 &        04-01 to 11-01 \\
\midrule
aggregate &         0.25 (0.21, 0.29) &        0.12 (0.08, 0.15) &  -0.54 (-0.69, -0.35) &     0.34 (0.32, 0.37) &         0.16 (0.14, 0.18) &   -0.53 (-0.6, -0.46) \\
30-39     &  - &  - &    - &  0.069 (0.063, 0.076) &       0.02 (0.013, 0.026) &   -0.72 (-0.81, -0.6) \\
40-49     & - &  - &  - &     0.12 (0.11, 0.12) &      0.032 (0.026, 0.038) &  -0.73 (-0.78, -0.67) \\
50-59     &  - &  - &  - &      0.19 (0.18, 0.2) &      0.067 (0.056, 0.077) &  -0.65 (-0.71, -0.59) \\
60-69     &         0.19 (0.15, 0.23) &      0.097 (0.054, 0.14) &  -0.49 (-0.74, -0.17) &      0.3 (0.29, 0.32) &         0.13 (0.11, 0.14) &  -0.58 (-0.63, -0.52) \\
70-79     &         0.33 (0.28, 0.37) &         0.15 (0.11, 0.2) &  -0.53 (-0.68, -0.36) &     0.44 (0.42, 0.46) &          0.2 (0.18, 0.22) &   -0.55 (-0.6, -0.49) \\
80+       &         0.47 (0.42, 0.52) &        0.27 (0.23, 0.31) &  -0.42 (-0.53, -0.29) &     0.61 (0.58, 0.62) &         0.33 (0.31, 0.35) &  -0.45 (-0.49, -0.41) \\
\bottomrule
\end{tabular}
  \end{minipage}}
\caption{Estimates of HFR and drop in HFR between April 1st and November 1st.
Median and $95\%$ confidence intervals are computed using block bootstrapping. Results with inadequate support are omitted.}
    \label{tab:hfr_drop_overall}
\end{table}

\section{Discussion}
\label{sec:discussion}
In this paper, we unpack the apparent improvement in fatality rates 
to hone in on improvements that could reasonably be attributed to advances in treatment. 
We account for shifting age distributions (H1) by age-stratifying, 
increased testing capacity (H2) by 
focusing on the hospitalized,
and the delay between detection and fatality (H3) 
by conducting a cohort-based analysis. 
We demonstrate that increased testing 
does not explain away the first and second peaks 
due to corresponding peaks in hospitalizations, 
deaths, and test positivity rates 
(Figures \ref{fig:pos_test_rate} and \ref{fig:covid_trend}). 
We visualize the shifting age distributions 
in cases, hospitalizations, and deaths over time 
(Figure \ref{fig:covid_ages}),
and we quantify the decrease in age-stratified HFRs 
between the two peaks (April 15th and July 15th), 
and between April 1st and December 1st. 
Putting all of these analyses together, 
we arrive at the following narrative: 

At the beginning of April, testing was relatively sparse 
(Figure \ref{fig:pos_test_rate}). 
Cases, hospitalizations, and deaths were rising, 
and reached peak levels on April 15th (Figure \ref{fig:covid_trend}). 
Roughly one in every ten tests was coming back positive in Florida, 
and one in every five tests was coming back positive nationally. 
In Florida, the aggregate HFR was approximately $24\%$,
with age-stratified HFRs ranging between
$9.2\%$ for the 50-59 age group to $47\%$ for the 80+ age group (Table \ref{tab:hfr_drop_peak}). 
Across the country, the aggegate HFR was at approximately $30\%$, 
while the age-stratified HFRs ranged between $5.5\%$ for the 30-39 age group 
and $57\%$ for the 80+ age group 
(Table \ref{tab:hfr_drop_peak}).
In each age group, the national HFR was higher than the Florida HFR, which
could be due to overwhelmed hospital systems in states
which were particularly hard hit during the first wave (e.g. New York). 
In fact, $48\%$ of national CDC cases 
between April 1st and April 15th 
were recorded in New York alone 
(Appendix Figure \ref{fig:top_5_states}). 

Over the next three months,
the proportion of younger individuals 
with COVID-19 grew steadily 
(Figure \ref{fig:covid_ages}). 
Testing continued to ramp up across the nation, 
and spiked in Florida as it approached 
a much heavier second peak around July 15 
(Figure \ref{fig:pos_test_rate}). 
Note, however that the corresponding positive test rates 
were also at an all-time high.
Florida experienced record hospitalizations and deaths, 
and the age-stratified HFR was every bit as high as in the first wave, 
perhaps even higher (Table \ref{tab:hfr_drop_peak}). 
While Bill Gates had publicly argued 
that due to improvements in treatment 
attributable to dexamethosone and remdesivir, 
``We've had a factor-of-two improvement in hospital outcomes already," \citep{levy2020gates} this did not yet appear to be the case in Florida. 
(Alternatively, it is also possible 
that treatment improvements might have 
been perfectly counterbalanced 
by the challenges of peak demand on the hospital system.) 
Nationally, on the other hand, cases in New York had diminished
(Appendix Figure \ref{fig:top_5_states}) and were starting to surge in other states, 
forming a smaller second peak 
(as measured by hospitalizations and deaths). 
Between the first peak and the second peak, 
the national HFR had dropped by $39\%$ in aggregate,
while the drop for age-stratified HFRs 
ranged between $50\%$ in the 59- age group 
and $24\%$ in the 80+ age group. 
The different stories told here by Florida 
and the national aggregate data underscore
the importance of state-level rather than national analysis.

Finally, come November 1st, age-stratified HFRs 
in both Florida and the national aggregate data 
appear to have dropped significantly, 
likely indicating treatment improvements 
(though possibly confounded 
by disease mutations (H5) 
and reduced viral loads (H6)).
We observe that in the national data, 
an increase in age corresponds
to a small relative drop in HFR.
Thus insofar as postulated improvements in treatment
might guide changes in policy, 
the comparatively small benefits 
in the elderly population should be taken into account. (Note however that the younger groups have small HFRs to begin with, so the opposite trend may appear when considering \emph{absolute} improvements rather than relative improvements.)
Since April 1st, the age-stratified HFR in Florida 
had decreased by as much as $53\%$ in the 70-79 age group 
and as little as $42\%$ in the 80+ age group. 
In the national data, 
the drop in age-stratified HFR 
was as much as $73\%$ in the 40-49 age group 
and as little as $45\%$ in the 80+ age group.
In reference to the case fatality ratio
\citep{factcheck_mcdonald_2020}, 
on July 27, President
Donald Trump stated 
in a press briefing that 
``Due to the medical advances we’ve already achieved 
and our increased knowledge in how to treat the virus, 
the mortality rate for patients over the age of 18 
is 85 percent lower than it was in April" \citep{trump_statement}.
Note, however, that none of our findings in Florida or nationally
are as large as the $85\%$ touted by President Trump 
(and re-running the analysis on patients 18+ does not change this). 
This emphasizes how the aggregate CFR 
can be misleading if age distribution shift, 
increased testing, and delays between detection 
and fatality are unaccounted for.

As far as we are aware, our analysis is first 
to explicitly account for age distribution shift, 
increased testing, and delays between detection and fatality. 
We recommend that policy makers account 
for at least these three factors, 
and show how in the absence of these considerations 
it is easy to be misled. 
More broadly, we advocate for more reliable, 
age-stratified hospitalization data from each state.
This would paint a much clearer picture 
when assessing the state of treatment improvements, 
and better inform both hospitals and policy-makers.

\section{Limitations}
\label{sec:limitations}
We aim to quantify treatment improvements (H4) 
in Florida and the United States 
by estimating changes in the age-stratified HFR. 
However, changes in the age-stratified HFR 
could also be influenced by disease mutation (H5) 
and changing viral loads (H6). 
To distinguish their effects in future work, 
we would need additional data quantifying these factors. 
Furthermore, while we listed the six main hypotheses
we found in our literature search of the decrease in case fatality rate, 
it is possible that alternative explanations may arise in the future, 
and those may need to be accounted for as well.

Additionally, we assume that treatment improvements 
will be reflected in the HFR because over our study's time range, 
major treatment improvements (e.g. dexamethosone and remdesivir)
targeted hospitalized patients. 
For future studies, we note that monoclonal antibody therapies bamlanivimab and the combination of casirivimab and imdevimab were recently authorized for emergency use 
in November \citep{monoclonal_eua, regeneron_monoclonal_eua}, 
and unlike dexamethosone and remdesivir, 
they are \emph{not} intended for hospitalized patients 
and instead recommended for patients 
likely to progress to severe COVID-19 and/or hospitalization. 
Thus, its effects may not be directly reflected in the HFR.

There are also several limitations which arise from data quality.
In both the Florida FDOH data and national CDC data,
missingness for hospitalization and death are high (Table \ref{tab:hosp_death}),
potentially introducing bias in the estimates for HFR 
if the data are not missing at random.
To make matters worse for the national CDC data, 
state-level plots of cases seem to indicate 
that each state may have different patterns 
of reporting their data to the CDC. 
First, the reported CDC cases appear
to be incomplete for several states.
In the subset of CDC data reported from Florida, 
the counts only account for $69\%$ of the cases 
provided by the Florida Department of Health, 
and even after 7-day smoothing,
the cases appear to be reported sporadically 
(Appendix Figure \ref{fig:covid_trend_aggregate}). 
Cross-referencing with the COVIDcast API\footnote{Other 
than the CDC data, we only have line level data 
from the Florida Department of Health.
Thus, for Texas, we cross-reference with 
aggregate case counts within our time range 
from the COVIDcast API, which uses data from USAFacts.}, 
we find that in the subset of CDC data from Texas, 
only $4.7\%$ of the cases 
and only $0.009\%$ of the deaths 
are accounted for \citep{covidcast2020api}. 
In fact, in the subset of the CDC data from Texas,
only $11$ hospitalizations were recorded 
across the entire studied time range.
In addition to missing cases, hospitalizations, and deaths, 
the CDC dataset has $63.5\%$ missingness 
for the positive specimen received date 
and $49.0\%$ missingness for the symptom onset date. 
While using the positive specimen received date 
would make the dates more comparable to those in Florida, 
we observe that more than half of the states 
have greater than $90\%$ missingness 
in the positive specimen date field 
(in fact in Florida it is $99.9\%$ missing). 
Due to these high levels of missingness not at random, 
we chose to use the CDC report date 
which is reported for all rows.

While the CDC report date does not have missing values, 
the daily cases based on CDC report date 
have spikes at certain days, 
which might be indicative of the reporting agency 
submitting several cases to CDC on the same day 
rather than reporting daily. 
While we excluded data from 
the states with the most extreme artifacts
(Appendix Figure \ref{fig:exclude_states}), 
there are still spikes in the remaining states 
(Appendix Figure \ref{fig:cdc_top12}). 
Thus, in our national analysis,
we rely on the hope that in aggregate,
the signal will outweigh the noise. 
Despite the data limitations, 
the CDC data appears to be the best available source 
of line-level cases needed for cohort-based analysis across the United States. 
We note that in the Florida FDOH data, on the other hand, 
we use the positive test confirmed date 
which is not missing at all in this data, 
making the Florida estimates 
more reliable 
than those from the national data.

\section*{Acknowledgments}
We thank Professors Roni Rosenfeld, Cosma Shalizi, and Marc Lipsitch 
for their detailed feedback 
throughout this analysis 
and the drafting of this manuscript. 

\bibliographystyle{plainnat}
\bibliography{refs}

\newpage
\appendix
\section{Aggregate plots}

\begin{figure}[ht!]
\centering
\begin{subfigure}[t]{0.93\textwidth}
\includegraphics[width=\textwidth]{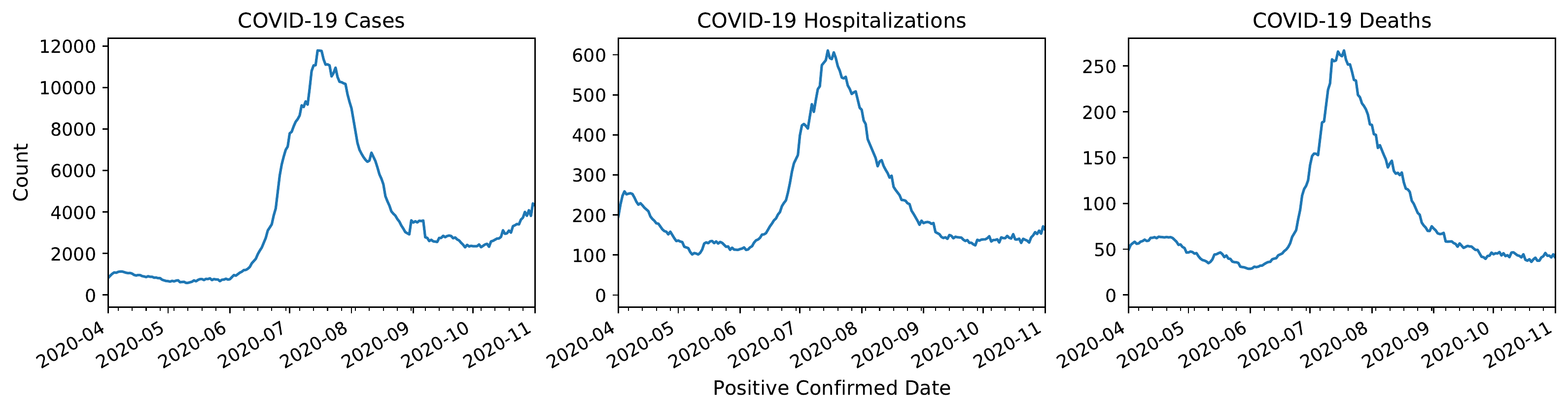}
\caption{Florida FDOH data. The $x$ axis is the date of positive test confirmation.}
\label{fig:covid_trend_fl}
\end{subfigure}
\hspace{0.5em}
\begin{subfigure}[t]{0.93\textwidth}
\vspace{0.8em}
\includegraphics[width=\textwidth]{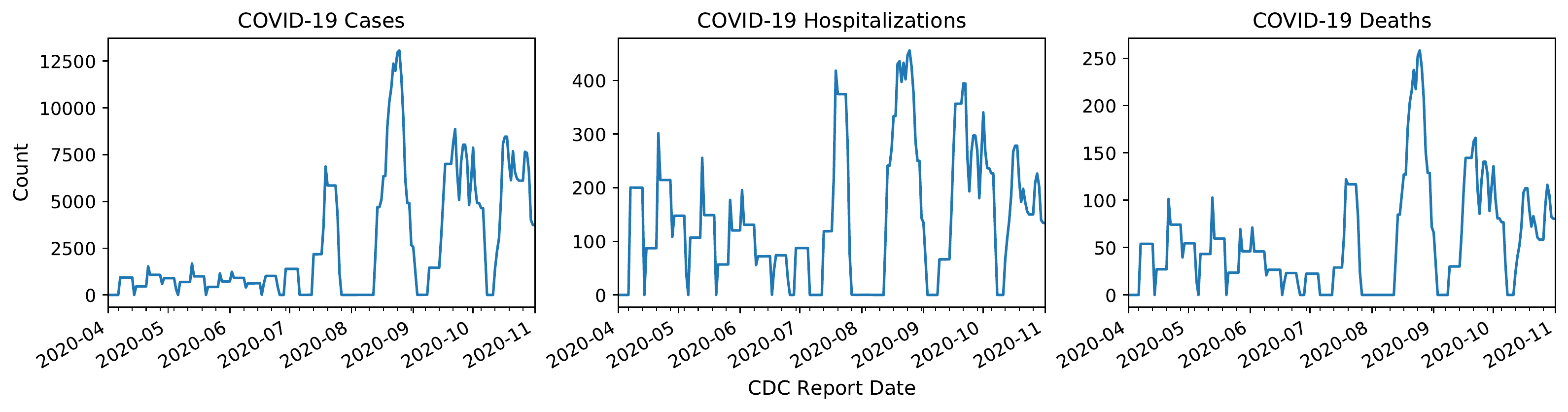}
\caption{Subset of United States CDC data in which state identifier is Florida. The positive specimen date for Florida rows is $99.99\%$ missing, so we must compare using the CDC report date.}
\label{fig:covid_trend_cdc_fl}
\end{subfigure}
\caption{Florida aggregate cases, (eventual) hospitalizations, and (eventual) deaths from FDOH data versus from United States CDC data.}
\label{fig:covid_trend_aggregate}
\end{figure}

\begin{figure}[H]
\centering
\includegraphics[width=0.93\textwidth]{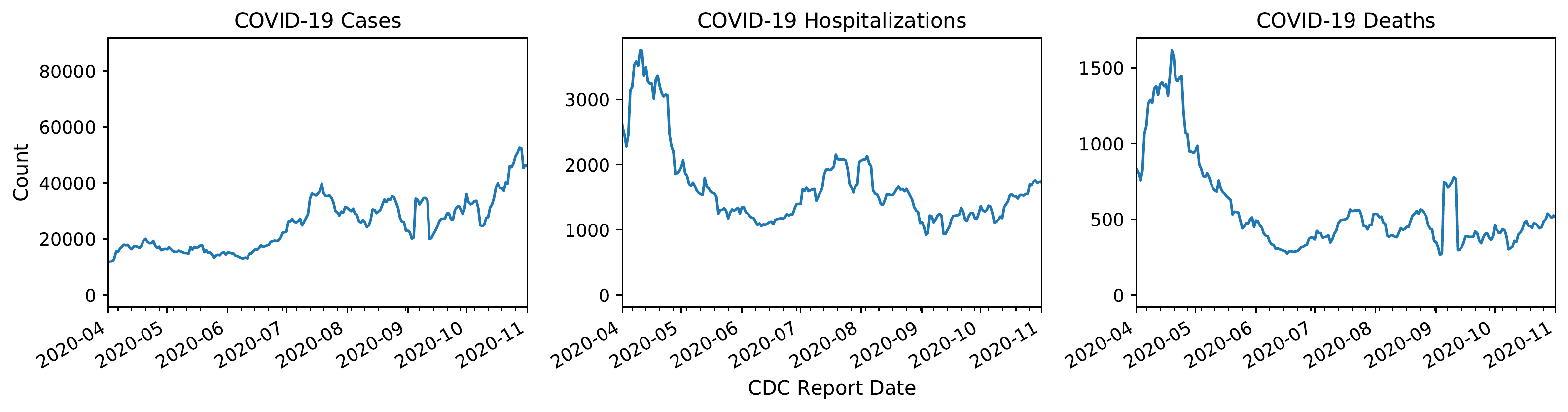}
\caption{Aggregate cases, (eventual) hospitalizations, and (eventual) deaths in country
by the date of report to the CDC (U.S.).}
\label{fig:country_covid_trend_agg}
\end{figure}

\section{Additional age-stratified plots}

\begin{figure}[H]
\centering
\begin{subfigure}[t]{\textwidth}
\includegraphics[width=\textwidth]{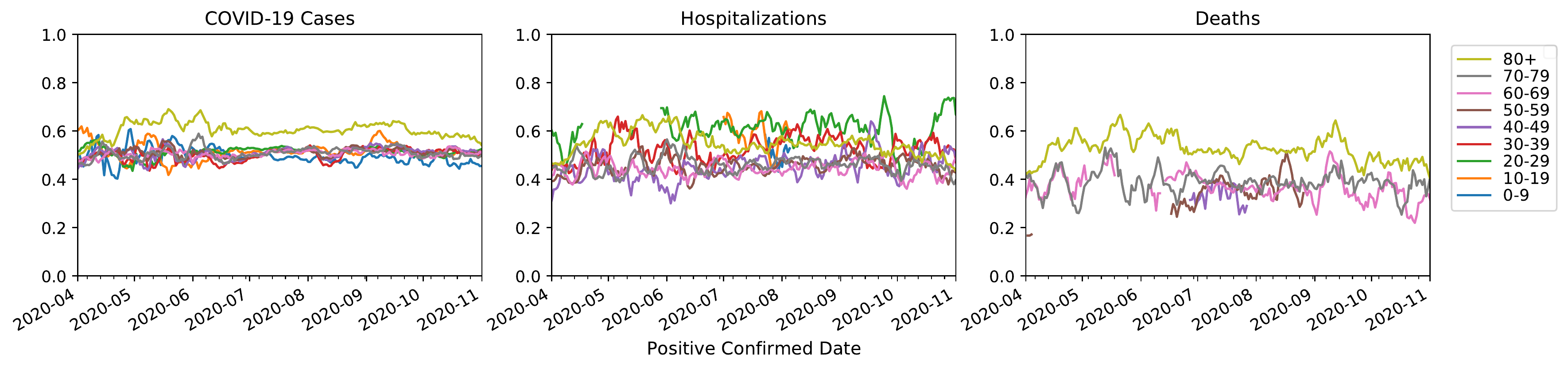}
\caption{Florida FDOH data}
\label{fig:florida_gender}
\end{subfigure}
\hspace{0.5em}
\begin{subfigure}[t]{\textwidth}
\vspace{0.8em}
\includegraphics[width=\textwidth]{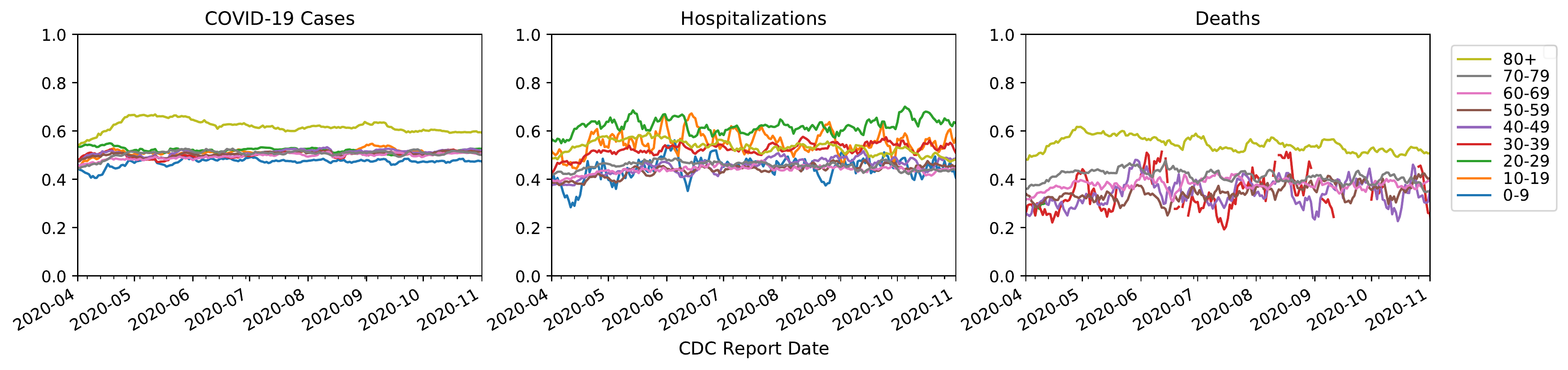}
\caption{United States CDC data}
\label{fig:country_gender}
\end{subfigure}
\caption{Female fraction of Florida and national cases, (eventual) hospitalizations, and (eventual) deaths vs. date of first positive test result (Florida) and date of report to the CDC (U.S.). Dates with fewer than five cases, hospitalizations, or deaths in the denominator are excluded.}
\label{fig:covid_gender}
\end{figure}

\begin{figure}[H]
\centering
\begin{subfigure}[t]{0.8\textwidth}
\includegraphics[width=\textwidth]{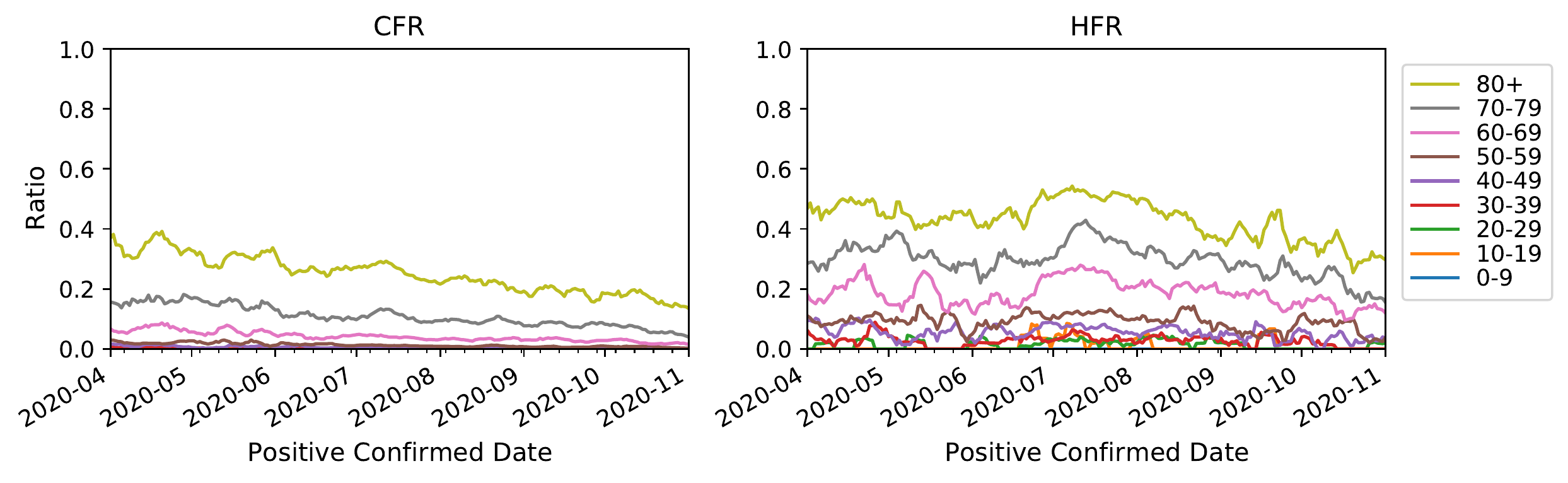}
\caption{Florida}
\label{fig:cfr_hfr_age}
\end{subfigure}
\hspace{0.5em}
\begin{subfigure}[t]{0.8\textwidth}
\vspace{0.8em}
\includegraphics[width=\textwidth]{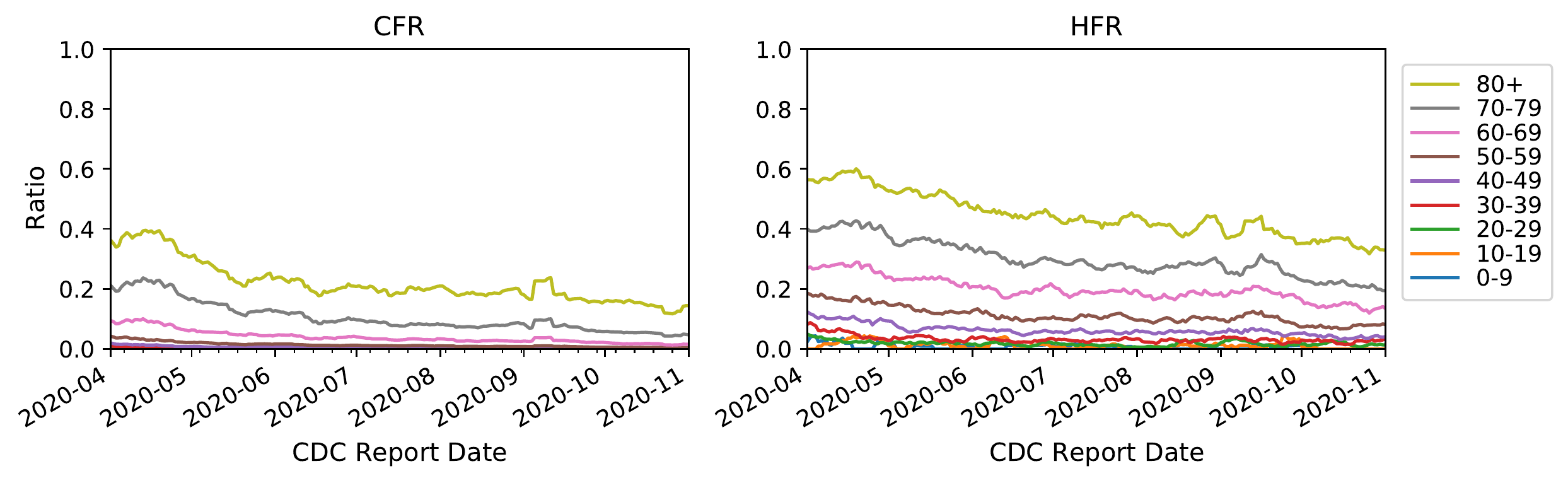}
\caption{National}
\label{fig:cfr_hfr_agg}
\end{subfigure}
\caption{Age-stratified CFR and HFR vs. date of first positive test result (Florida) and date of report to the CDC (U.S.)}
\label{fig:cfr_hfr_ratio}
\end{figure}

\section{State-wise plots}
\begin{figure}[H]
\centering
\includegraphics[width=\textwidth]{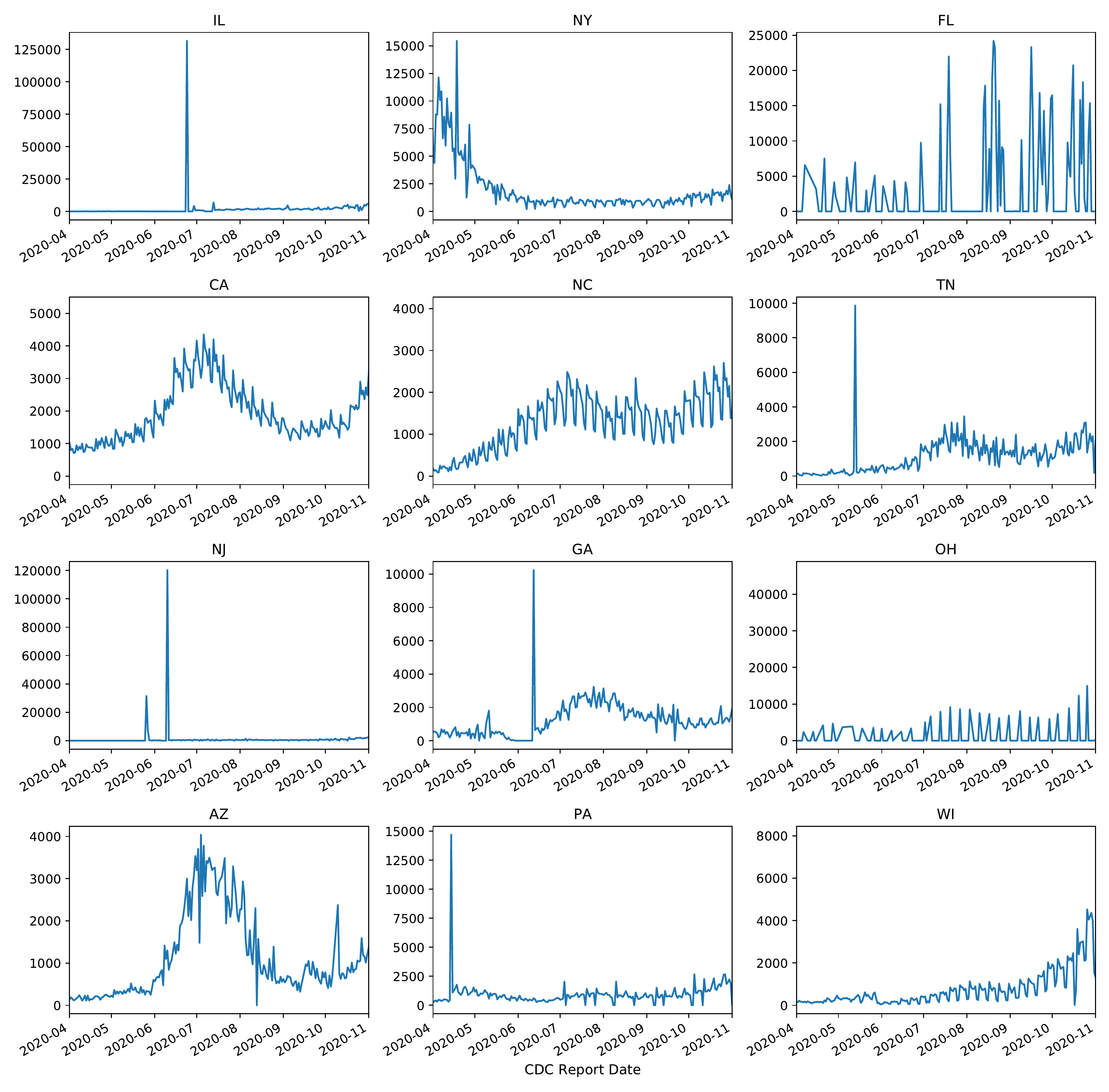}
\caption{Daily lab-confirmed COVID-19 cases in top twelve states with most COVID-19 cases from United States CDC data, by date of report to the CDC (U.S.). Note we graph daily cases instead of the 7-day average over cases in order to demonstrate the nature of the raw data.}
\label{fig:cdc_top12}
\end{figure}

\begin{figure}[H]
\centering
\includegraphics[width=\textwidth]{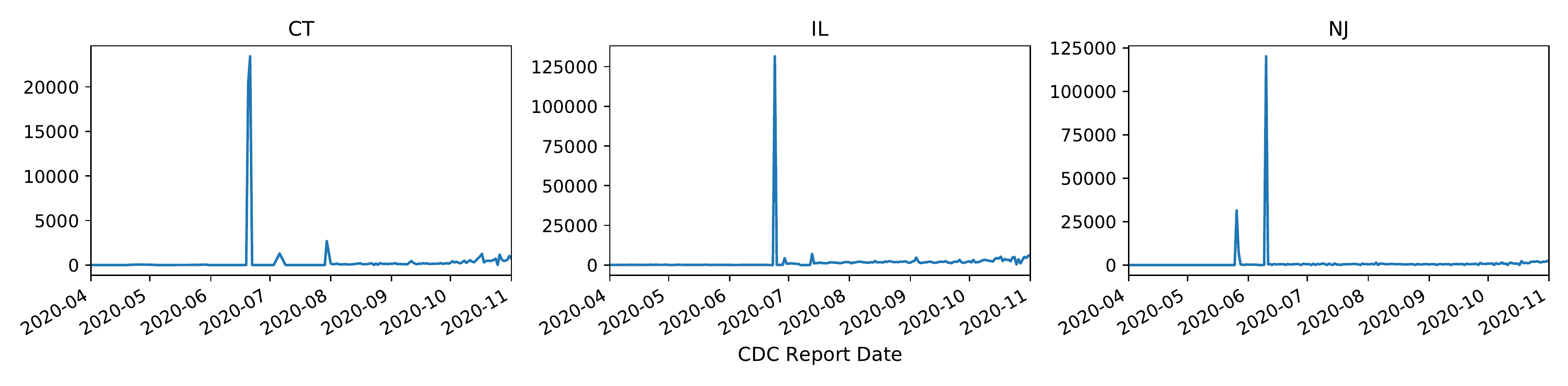}
\caption{Daily lab-confirmed COVID-19 cases in NJ, IL, and CT from United States CDC data vs. date of report to the CDC}
\label{fig:exclude_states}
\end{figure}

\begin{figure}[H]
\centering
\begin{subfigure}[t]{\textwidth}
\includegraphics[width=\textwidth]{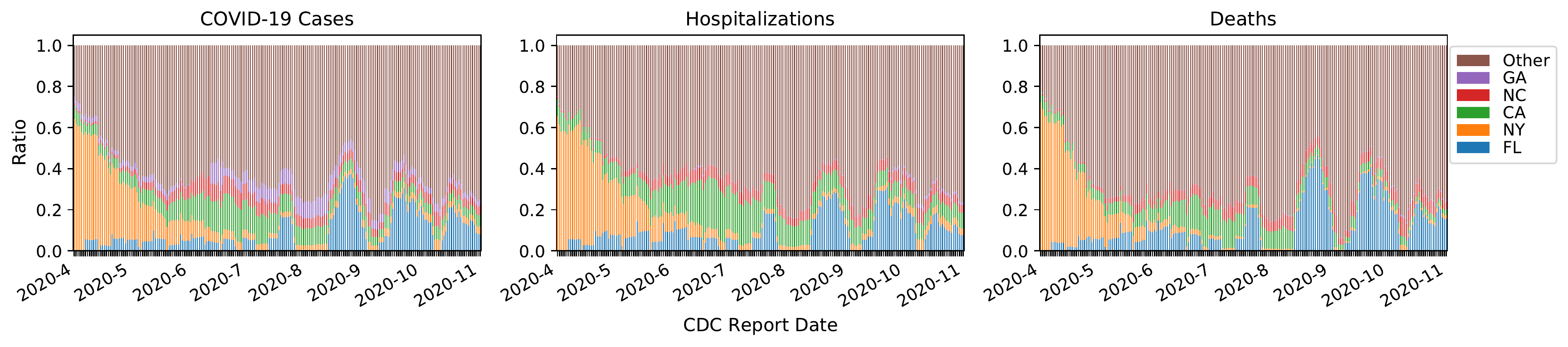}
\end{subfigure}
\caption{State-wise distribution of cases, 
(eventual) hospitalizations,
and (eventual) deaths for top five states FL, NY, CA, GA and NC (and the rest) with most COVID-19 cases from United States CDC data, by date of report to the CDC (U.S.)}
\label{fig:top_5_states}
\end{figure}

\section{Walkthrough of residual block-bootstrapping with post-blackening} \label{app:block_bootstrap}
To estimate the trend in HFR with uncertainty, we follow the steps below:
\begin{enumerate}
    \item For each age group, fit a smoothing spline (3rd order) to the 7-day lagged average HFR. This provides an estimate of the trend. Now we would like confidence intervals around this trend.
    \item Take residuals from fitting the cubic spline. Block-bootstrap sample 1000 replicates with 7-day block sizes. This gives us a dataset of residuals the same size as the original dataset.
    \item For each of the 1000 replicates, add the sample residuals onto the estimated trend in step 1. This is called ``post-blackening" \citep{bootstrap_davison_hinkley_1997}, and it gives us 1000 new datasets drawn from the same distribution as the original time series assuming uncorrelated blocks of residuals. 
    \item On each of the 1000 new datasets, re-estimate the trend using the smoothing spline. At every point in time, we can use the estimated trends from these 1000 replicates in order to get confidence intervals.
\end{enumerate}

\begin{figure}[H]
\centering
\begin{subfigure}[t]{0.8\textwidth}
\includegraphics[width=\textwidth]{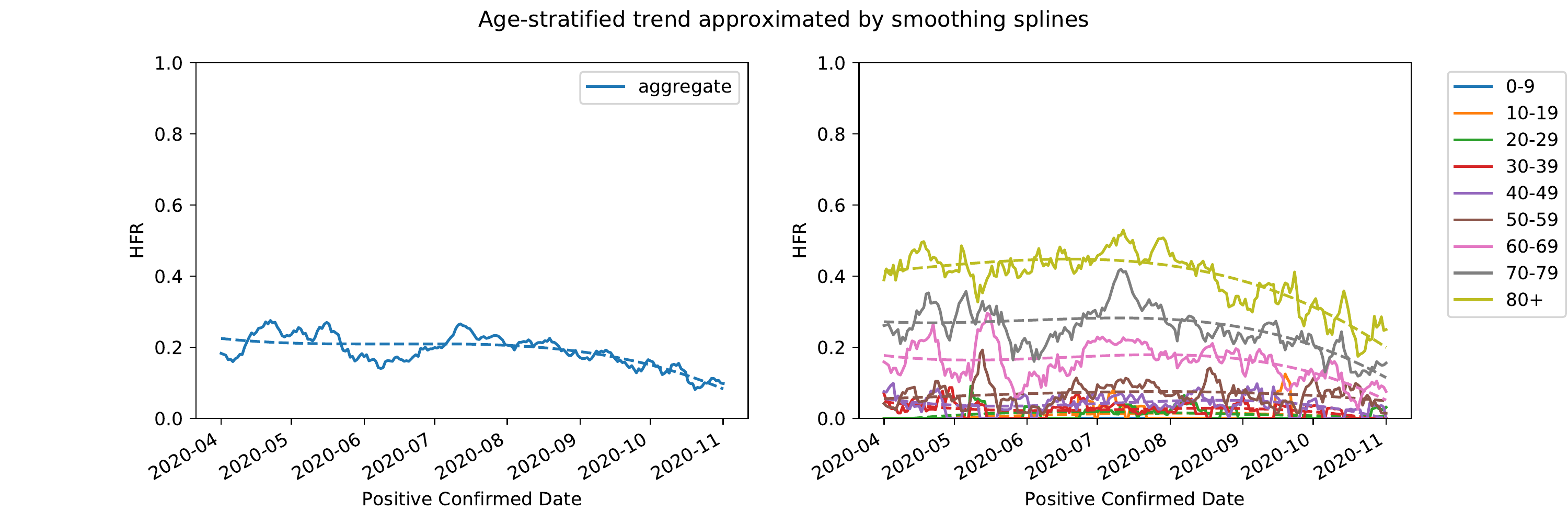}
\caption{Estimate of trend in age-stratified Florida HFRs, fit using cubic smoothing splines.}
\label{fig:cubic_spline_florida}
\end{subfigure}
\hspace{0.5em}
\begin{subfigure}[t]{0.8\textwidth}
\vspace{0.8em}
\includegraphics[width=\textwidth]{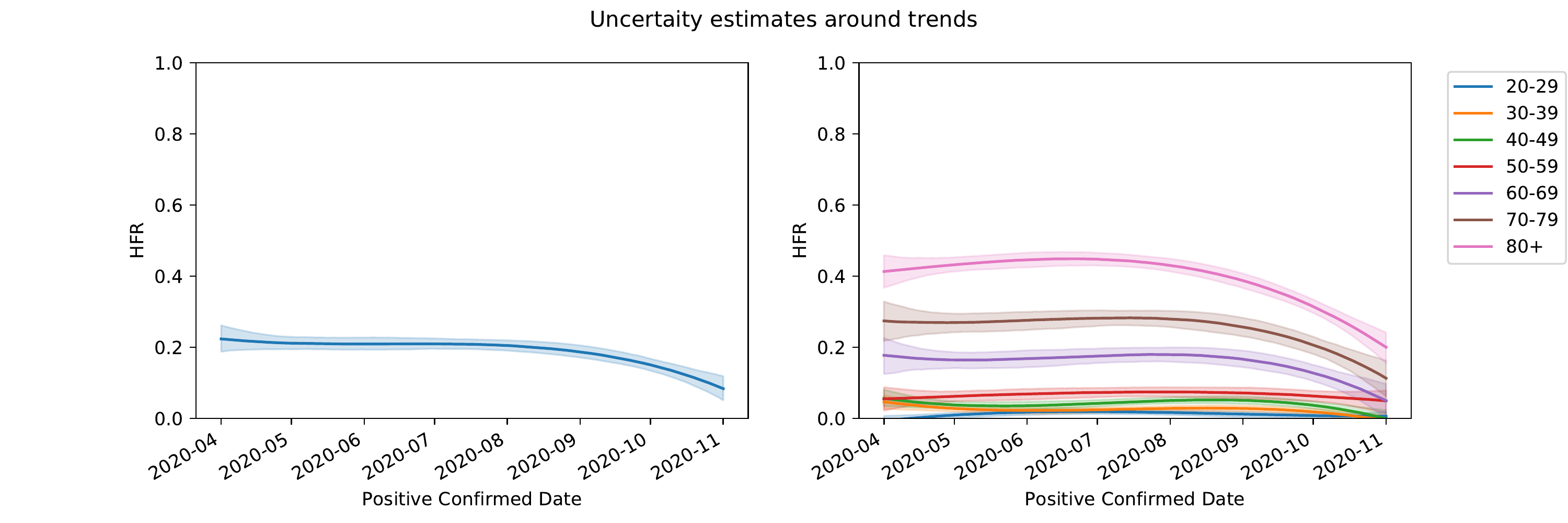}
\caption{Estimate of trend and uncertainty around age-stratified Florida HFRs, derived using residual block-bootstrapping with post-blackening.}
\label{fig:residual_bootstrap_florida}
\end{subfigure}
\caption{Estimate of trend in age-stratified HFRs in Florida.}
\label{fig:walkthough_florida}
\end{figure}

\begin{figure}[H]
\centering
\begin{subfigure}[t]{0.8\textwidth}
\includegraphics[width=\textwidth]{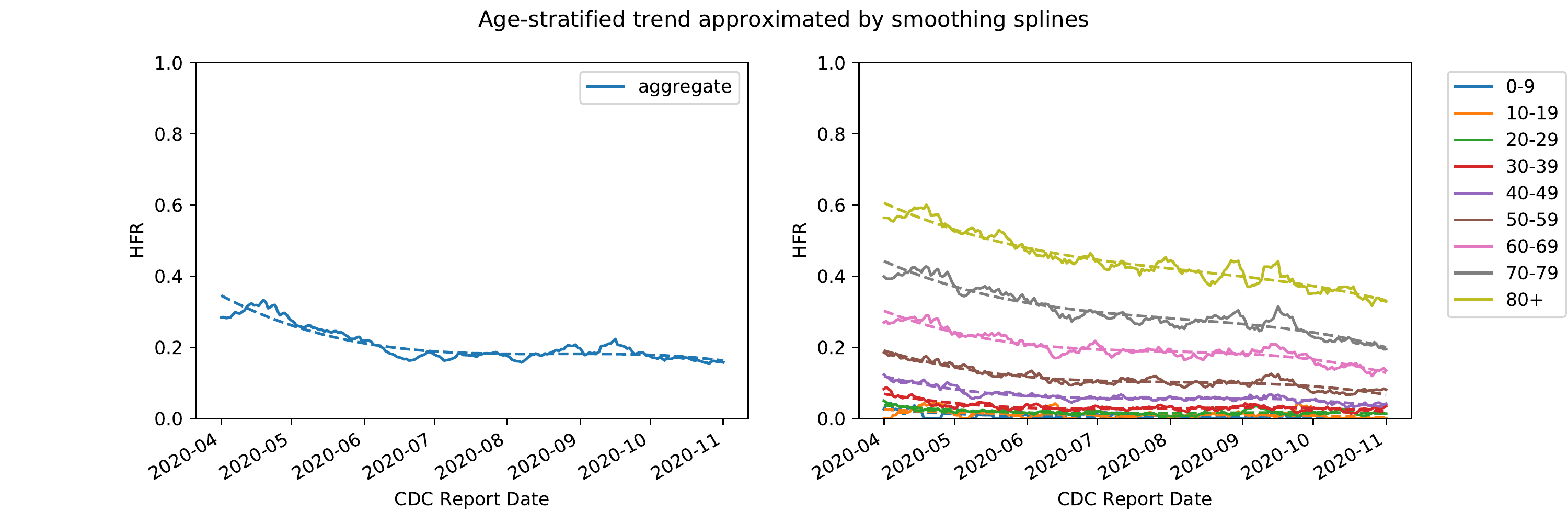}
\caption{Estimate of trend in age-stratified national HFRs, fit using cubic smoothing splines.}
\label{fig:cubic_spline_national}
\end{subfigure}
\hspace{0.5em}
\begin{subfigure}[t]{0.8\textwidth}
\vspace{0.8em}
\includegraphics[width=\textwidth]{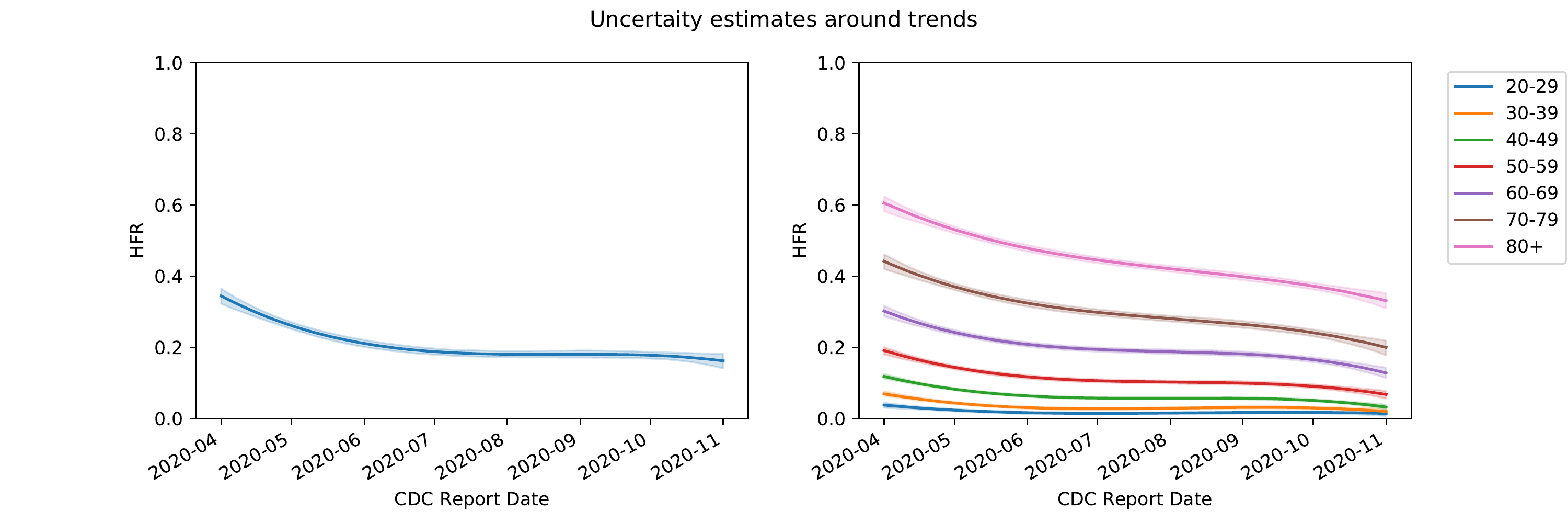}
\caption{Estimate of trend and uncertainty around age-stratified national HFRs, derived using residual block-bootstrapping with post-blackening.}
\label{fig:residual_bootstrap_national}
\end{subfigure}
\caption{Estimate of trend in age-stratified HFRs in national CDC data.}
\label{fig:walkthrough_national}
\end{figure}

Overall, the cubic smoothing splines appear to fit the 7-day lagged average HFRs relatively well (Figures \ref{fig:cubic_spline_florida} and \ref{fig:cubic_spline_national}). In Florida, the estimates of HFR trend and uncertainty appear to show a relatively flat trend for most of the time range, with a recent decline in HFR in the last two months (Figure \ref{fig:residual_bootstrap_florida}). Nationally, the estimates of HFR trend and uncertainty appear to show a more consistent decline in HFR over the entire time range (Figure \ref{fig:residual_bootstrap_national}).

\section{Age-stratified drops in HFR by gender}\label{app:gender_hfr_drop}
\begin{table}[H]
      \setlength{\tabcolsep}{0.5em}
  \resizebox{0.8\textwidth}{!}{\begin{minipage}{\textwidth}
    \centering
\begin{tabular}{l|ccc|ccc}
\toprule
& \multicolumn{3}{c|}{\textbf{Florida}} & \multicolumn{3}{c}{\textbf{National}} \\
\hline
\textbf{Age group}& 2020-04-15 &            2020-07-15 &        04-15 to 07-15 &            2020-04-15 &             2020-07-15 &        04-15 to 07-15 \\
\midrule
aggregate &      0.22 (0.19, 0.24) &     0.21 (0.19, 0.22) &   -0.034 (-0.16, 0.1) &     0.28 (0.26, 0.29) &      0.16 (0.15, 0.17) &  -0.43 (-0.47, -0.38) \\
30-39     &   -&-&- &  -&-&- \\
40-49     &   -&-&- &  0.075 (0.068, 0.081) &    0.046 (0.042, 0.05) &  -0.38 (-0.46, -0.29) \\
50-59     &   -&-&- &     0.14 (0.13, 0.14) &    0.08 (0.075, 0.085) &  -0.41 (-0.46, -0.36) \\
60-69     &       0.17 (0.14, 0.2) &      0.18 (0.16, 0.2) &   0.062 (-0.15, 0.36) &     0.23 (0.22, 0.24) &      0.16 (0.16, 0.17) &  -0.29 (-0.33, -0.25) \\
70-79     &       0.27 (0.23, 0.3) &      0.28 (0.26, 0.3) &    0.046 (-0.1, 0.24) &     0.36 (0.35, 0.37) &      0.25 (0.25, 0.26) &   -0.3 (-0.33, -0.26) \\
80+       &       0.42 (0.4, 0.45) &     0.44 (0.43, 0.46) &  0.047 (-0.035, 0.14) &     0.53 (0.52, 0.55) &       0.39 (0.38, 0.4) &  -0.26 (-0.29, -0.23) \\
\bottomrule
\end{tabular}
  \end{minipage}}
\caption{Estimates of HFR and drop in HFR on peak dates and among females. Median and $95\%$ confidence intervals are computed using block bootstrapping.}
\end{table}

\begin{table}[H]
      \setlength{\tabcolsep}{0.5em}
  \resizebox{0.8\textwidth}{!}{\begin{minipage}{\textwidth}
    \centering
\begin{tabular}{l|ccc|ccc}
\toprule
& \multicolumn{3}{c|}{\textbf{Florida}} & \multicolumn{3}{c}{\textbf{National}} \\
\hline
\textbf{Age group}&            2020-04-15 &            2020-07-15 &          04-15 to 07-15 &            2020-04-15 &            2020-07-15 &        04-15 to 07-15 \\
\midrule
aggregate &      0.25 (0.23, 0.28) &     0.25 (0.24, 0.27) &   -0.0077 (-0.13, 0.13) &     0.32 (0.31, 0.33) &       0.2 (0.2, 0.21) &   -0.36 (-0.4, -0.32) \\
30-39     &   -&-&- &  0.074 (0.066, 0.083) &  0.036 (0.031, 0.041) &  -0.52 (-0.61, -0.42) \\
40-49     &   -&-&- &     0.12 (0.11, 0.12) &   0.066 (0.062, 0.07) &  -0.43 (-0.47, -0.38) \\
50-59     &  0.12 (0.1, 0.14) &     0.13 (0.12, 0.14) &     0.069 (-0.11, 0.29) &     0.19 (0.18, 0.19) &     0.12 (0.12, 0.13) &   -0.34 (-0.38, -0.3) \\
60-69     &       0.2 (0.17, 0.23) &     0.23 (0.22, 0.25) &      0.17 (-0.01, 0.43) &      0.3 (0.29, 0.31) &     0.21 (0.21, 0.22) &  -0.29 (-0.32, -0.25) \\
70-79     &      0.36 (0.33, 0.39) &     0.37 (0.35, 0.39) &    0.029 (-0.065, 0.14) &     0.44 (0.42, 0.46) &     0.32 (0.31, 0.33) &   -0.27 (-0.3, -0.23) \\
80+       &      0.52 (0.49, 0.56) &      0.52 (0.5, 0.55) &  0.0013 (-0.084, 0.088) &      0.61 (0.6, 0.62) &     0.48 (0.47, 0.49) &  -0.21 (-0.23, -0.18) \\
\bottomrule
\end{tabular}

  \end{minipage}}
\caption{Estimates of HFR and drop in HFR on peak dates and among males. Median and $95\%$ confidence intervals are computed using block bootstrapping.}
\end{table}

\begin{table}[H]
      \setlength{\tabcolsep}{0.5em}
  \resizebox{0.8\textwidth}{!}{\begin{minipage}{\textwidth}
    \centering
\begin{tabular}{l|ccc|ccc}
\toprule
& \multicolumn{3}{c|}{\textbf{Florida}} & \multicolumn{3}{c}{\textbf{National}} \\
\hline
\textbf{Age group}&            2020-04-01 &                2020-11-01 &        04-01 to 11-01 &            2020-04-01 &              2020-11-01 &        04-01 to 11-01 \\
\midrule
aggregate &         0.22 (0.19, 0.26) &       0.084 (0.051, 0.12) &  -0.63 (-0.78, -0.43) &      0.32 (0.3, 0.34) &       0.14 (0.12, 0.16) &  -0.56 (-0.64, -0.48) \\
30-39     &   -&-&- &  -&-&- \\
40-49     &      -&-&- &  0.085 (0.074, 0.095) &     0.032 (0.021, 0.04) &  -0.63 (-0.76, -0.49) \\
50-59     &      -&-&- &     0.16 (0.15, 0.17) &    0.055 (0.043, 0.066) &  -0.66 (-0.74, -0.58) \\
60-69     &      -&-&- &     0.25 (0.24, 0.27) &        0.12 (0.1, 0.13) &   -0.54 (-0.6, -0.46) \\
70-79     &         0.27 (0.22, 0.33) &        0.11 (0.062, 0.16) &  -0.59 (-0.79, -0.33) &     0.39 (0.37, 0.41) &       0.17 (0.15, 0.19) &  -0.56 (-0.62, -0.49) \\
80+       &         0.41 (0.37, 0.46) &          0.2 (0.16, 0.24) &  -0.52 (-0.64, -0.37) &      0.58 (0.55, 0.6) &       0.29 (0.26, 0.31) &   -0.5 (-0.55, -0.45) \\
\bottomrule
\end{tabular}
  \end{minipage}}
\caption{Estimates of HFR and drop in HFR between April 1st and November 1st and among females. Median and $95\%$ confidence intervals are computed using block bootstrapping.}
\end{table}

\begin{table}[H]
      \setlength{\tabcolsep}{0.5em}
  \resizebox{0.8\textwidth}{!}{\begin{minipage}{\textwidth}
    \centering
\begin{tabular}{l|ccc|ccc}
\toprule
& \multicolumn{3}{c|}{\textbf{Florida}} & \multicolumn{3}{c}{\textbf{National}} \\
\hline
\textbf{Age group}&            2020-04-01 &                2020-11-01 &        04-01 to 11-01 &            2020-04-01 &            2020-11-01 &        04-01 to 11-01 \\
\midrule
aggregate &      0.27 (0.23, 0.32) &          0.14 (0.1, 0.18) &  -0.47 (-0.64, -0.28) &     0.37 (0.34, 0.39) &      0.18 (0.16, 0.2) &   -0.5 (-0.58, -0.43) \\
30-39     &    -&-&- &   0.094 (0.082, 0.11) &   0.03 (0.018, 0.043) &  -0.68 (-0.82, -0.51) \\
40-49     &     -&-&- &     0.14 (0.13, 0.15) &  0.033 (0.022, 0.042) &  -0.77 (-0.85, -0.69) \\
50-59     &     -&-&- &      0.21 (0.2, 0.22) &  0.077 (0.063, 0.088) &  -0.64 (-0.71, -0.57) \\
60-69     &       0.2 (0.15, 0.25) &        0.13 (0.082, 0.17) &  -0.35 (-0.63, 0.013) &     0.34 (0.32, 0.35) &     0.14 (0.12, 0.15) &  -0.59 (-0.65, -0.53) \\
70-79     &      0.37 (0.33, 0.42) &         0.19 (0.14, 0.23) &   -0.5 (-0.64, -0.34) &     0.48 (0.46, 0.51) &      0.22 (0.2, 0.25) &   -0.54 (-0.6, -0.47) \\
80+       &      0.53 (0.48, 0.59) &         0.35 (0.29, 0.41) &  -0.35 (-0.49, -0.19) &     0.64 (0.61, 0.66) &      0.38 (0.36, 0.4) &   -0.4 (-0.45, -0.36) \\
\bottomrule
\end{tabular}
  \end{minipage}}
\caption{Estimates of HFR and drop in HFR between April 1st and November 1st and among males. Median and $95\%$ confidence intervals are computed using block bootstrapping.}
\end{table}

\end{document}